\documentclass[fleqn,usenatbib]{mnras}

\usepackage{newtxtext,newtxmath}
\usepackage[T1]{fontenc}
\usepackage{ae,aecompl}

\usepackage{graphicx}	
\usepackage{amsmath}	
\usepackage{amssymb}	
\usepackage{booktabs,caption}
\usepackage[flushleft]{threeparttable}

\title[SN2019gsc]{Observations of the low-luminosity Type~Iax supernova 2019gsc: a fainter clone of SN~2008ha?\thanks{This paper includes data gathered  with the 2.56-m Nordic Optical Telescope, the 10.4-m Gran Telescopio Canarias and the Liverpool Telescope located at the Observatorio del Roque de los Muchachos, La Palma, Spain. It also includes data obtained with Las Cumbres Observatory global network. Data available on request.}}

\author[L. Tomasella et al.]{Lina Tomasella,$^{1}$\thanks{E-mail: lina.tomasella@inaf.it}
Maximilian Stritzinger,$^{2}$
Stefano Benetti,$^{1}$
Nancy Elias-Rosa,$^{1,3}$
\newauthor
Enrico Cappellaro,$^{1}$
Erkki Kankare,$^{4}$
Peter Lundqvist,$^{5,6}$
Mark Magee,$^{7}$
\newauthor
Kate Maguire,$^{7}$
Andrea Pastorello,$^{1}$
Simon Prentice,$^{7}$
Andrea Reguitti$^{1,8,9}$
\\
$^{1}$INAF Osservatorio Astronomico di Padova, Vicolo dell'Osservatorio 5, 35122 Padova, Italy\\
$^{2}$Department of Physics and Astronomy, Aarhus University, Ny Munkegade 120, DK-8000 Aarhus C, Denmark\\
$^{3}$Institute of Space Sciences (ICE, CSIC), Campus UAB, Carrer de Can Magrans s/n, 08193 Barcelona, Spain\\
$^{4}$Tuorla Observatory, Department of Physics and Astronomy, University of
Turku, FI-20014 Turku, Finland\\
$^{5}$Department of Astronomy, AlbaNova University Center, Stockholm University, SE-10691 Stockholm, Sweden\\
$^{6}$The Oskar Klein Centre, AlbaNova, SE-10691 Stockholm, Sweden\\
$^{7}$School of Physics, Trinity College Dublin, Dublin 2, Ireland\\
$^{8}$Millennium Institute of Astrophysics (MAS), Nuncio Monse\~nor Sotero Sanz 100, Providencia, Santiago, Chile \\
$^{9}$Departamento de Ciencias Fisicas, Universidad Andres Bello, Fernandez Concha 700, Las Condes, Santiago, Chile\\
}

\date{Accepted MNRAS}

\pubyear{2020}

\begin{document}
\label{firstpage}
\pagerange{\pageref{firstpage}--\pageref{lastpage}}
\maketitle
\begin{abstract}
We present optical photometric and spectroscopic observations of the faint-and-fast evolving type~Iax SN~2019gsc, extending from the time of $g$-band maximum until about fifty days post maximum, when the object faded to an apparent $r$-band magnitude $m_r = 22.48\pm 0.11$ mag. 
SN~2019gsc reached a peak luminosity of only $M_{g} = -13.58\pm0.15$ mag, and is characterised with a post-maximum decline rate $\Delta m_{15}(g) =  1.08\pm0.14$ mag. These light curve parameters are comparable to those measured for SN~2008ha of $M_g = -13.89\pm0.14$ mag at peak and $\Delta m_{15}$(g) $=1.80\pm0.03$ mag. The spectral features of SN~2019gsc also resemble those of SN~2008ha at similar phases. This includes both the extremely low ejecta velocity at maximum, $\sim$3,000~km~s$^{-1}$, and at late-time (phase $+$54~d) strong forbidden iron and cobalt lines as well as both forbidden and permitted calcium features.
Furthermore, akin to SN~2008ha, the bolometric light curve of SN~2019gsc is consistent with the production  of $\approx 0.003\pm0.001$~M$_{\odot}$ of $^{56}$Ni. The explosion parameters, $M_{\textit{ej}} \approx 0.13$ M$_{\odot}$ and $E_k \approx 12 \times  10^{48}$ erg, are also similar to those inferred for SN~2008ha.
We estimate a sub-solar oxygen abundance for the host galaxy of SN~2019gsc (12 + log$_{\rm 10}$(O/H) $= 8.10 \pm 0.18$ dex), consistent with the equally metal-poor environment of SN~2008ha. Altogether, our dataset for SN~2019gsc indicates that this is a member of a small but growing group of extreme SN~Iax that includes SN~2008ha and SN~2010ae.

\end{abstract}




\section{Introduction}

Type~Iax supernovae (hereafter SNe Iax), also known as 2002cx-like SNe after their prototype described by \citet{li2003}, are possibly a class of thermonuclear explosions showing some spectroscopic similarities to SNe~Ia near maximum light, while at late-time they appear significantly different  \citep[e.g.,][]{foley2016,jha2017}. In fact, at early epochs SNe Iax exhibit features of \ion{Fe}{iii} and therefore resemble over-luminous 1991T-like SNe at similar phases. However, their late spectra are dominated by prevalent \ion{Ca}{ii} permitted and forbidden spectral lines in complete contrast to SNe~Ia that are instead  dominated by broad, forbidden [\ion{Fe}{ii}] and [\ion{Fe}{iii}] emission features.
Interestingly, the energetics of SNe Iax explosions and their luminosities
can be very different from one event to the other: within the class, peak absolute magnitudes range from $M_{\textit{peak}} \approx -13$ mag for the fainter members to $M_{\textit{peak}} \approx -19$ mag for the brighter ones (the latter close to the peak luminosity of normal SNe~Ia; cf. \citealt{jha2017}, see his Table~1). In all cases, SNe~Iax show maximum-light expansion velocities measured from the \ion{Si}{II} $\lambda$6355 line that are much lower than those measured in SNe~Ia. 
Quantitatively, at maximum the expansion velocities in SNe~Ia are typically around $\sim 10,000$~km~s$^{-1}$, while the velocities inferred from SNe~Iax never exceed $7,000$~km~s$^{-1}$, and can be as as low as 2,000~km~s$^{-1}$. 

The above mentioned peculiar traits of SNe~Iax, particularly of the fainter members of the class such as SN~2008ha and SN~2010ae \citep{foley2009,valenti2009,stritzinger14}, have brought out a variety of models and explosion scenarios \citep[see the review by][]{jha2017}, including a core-collapse, massive star supernova origin \citep{moriya2010}. However, the bunch of the observational characteristics of SNe~Iax suggests their close affinity to normal type Ia SNe and to several other classes of peculiar SNe Ia \citep[cf.][]{taub2017}. Thus, the thermonuclear explosion of a carbon-oxygen Chandrasekhar-mass ($M_{\textit{Ch}}$) white dwarf (WD) in a binary system is the leading scenario that has emerged gradually over time \citep{foley2009,stritzinger14,stritzinger15,jha2017}.

In the context of the thermonuclear explosion of WDs, a number of alternatives have been proposed to account for the origins of SNe~Iax and to explain the diversity in luminosities and explosion energies among the members of the class, from \textit{failed} deflagrations that leave a bound remnant (which is a natural outcome of 3D models with off-centre ignition, cf. \citealt{nonaka2012}) to pulsational delayed detonations \citep[i.e. for the brightest SNe 2005hk or 2012Z, see ][]{stritzinger15}. The direct detection of a blue source at the location of SN~2012Z in pre-explosion \textit{Hubble Space Telescope} (HST) images by \citet{mccully2014} suggested a helium-star donating material to an accreting WD as the progenitor of this event, similarly to the Galactic helium nova V445-Pup  \citep{mccully2014}.  

Pure deflagration models of carbon-oxygen (C/O) WDs or hybrid carbon-oxygen-neon (C/O/Ne) WDs in binary systems (likely with a He-star companion), as developed by \cite{jordan2012,kromer2013,fink2014,kromer2015}, can replicate the overall properties of intermediate-to-bright SNe~Iax. However, some difficulties have arisen with the faintest members of the class, i.e. in reproducing the observed amount of nickel (it is too large in \citealt{fink2014} models), or ejecta mass (it is too small in \citealt{kromer2015} simulations). Thus, in order to test the various models proposed for SNe~Iax, including the merger of C/O$-$O/Ne WDs as in the double-degenerate scenario of \cite{kashyap2018}, a larger observational dataset of SNe~Iax is needed, especially of the low end of the peak luminosity distribution.

Soon after discovery, \citet{leloudas2019} provided the spectroscopic classification of SN 2019gsc as a type~Iax event. Furthermore, the peculiar nature of this object was noted, as it closely resembled the type~Iax SNe~2008ha and 2010ae \citep{foley2009,valenti2009,stritzinger14}, the faintest and least-energetic SNe Iax yet observed. The preliminary photometric observations of SN~2019gsc indicated that it could be even fainter than the extreme SN~2008ha. 
This paper presents optical spectroscopic and photometric observations of SN~2019gsc that, as we will show, shares many characteristics with both SN~2008ha and SN~2010ae over its entire evolution. The manuscript is organized as follows: in Section~2 we provide basic information about the SN discovery, its host galaxy, our observations and the data reduction procedures. In Section~3, the photometric evolution and photometric parameters of SN~2019gsc are presented, while the spectroscopic analysis is reported in Section~4. In Section~5 we examine the environment of SN~2019gsc and set it in context with that of other faint SNe~Iax. Discussion and conclusions are in Section~6. 
Another paper on SN~2019gsc by \cite{srivastav2020} was submitted for publication almost simultaneously to this work. A brief comparison of the results of these independent investigations is also reported in the final section.

\section{Discovery and follow-up observations of SN~2019gsc}
\label{discovery}

\begin{figure}
	\includegraphics[width=\columnwidth, angle=0, scale=1]{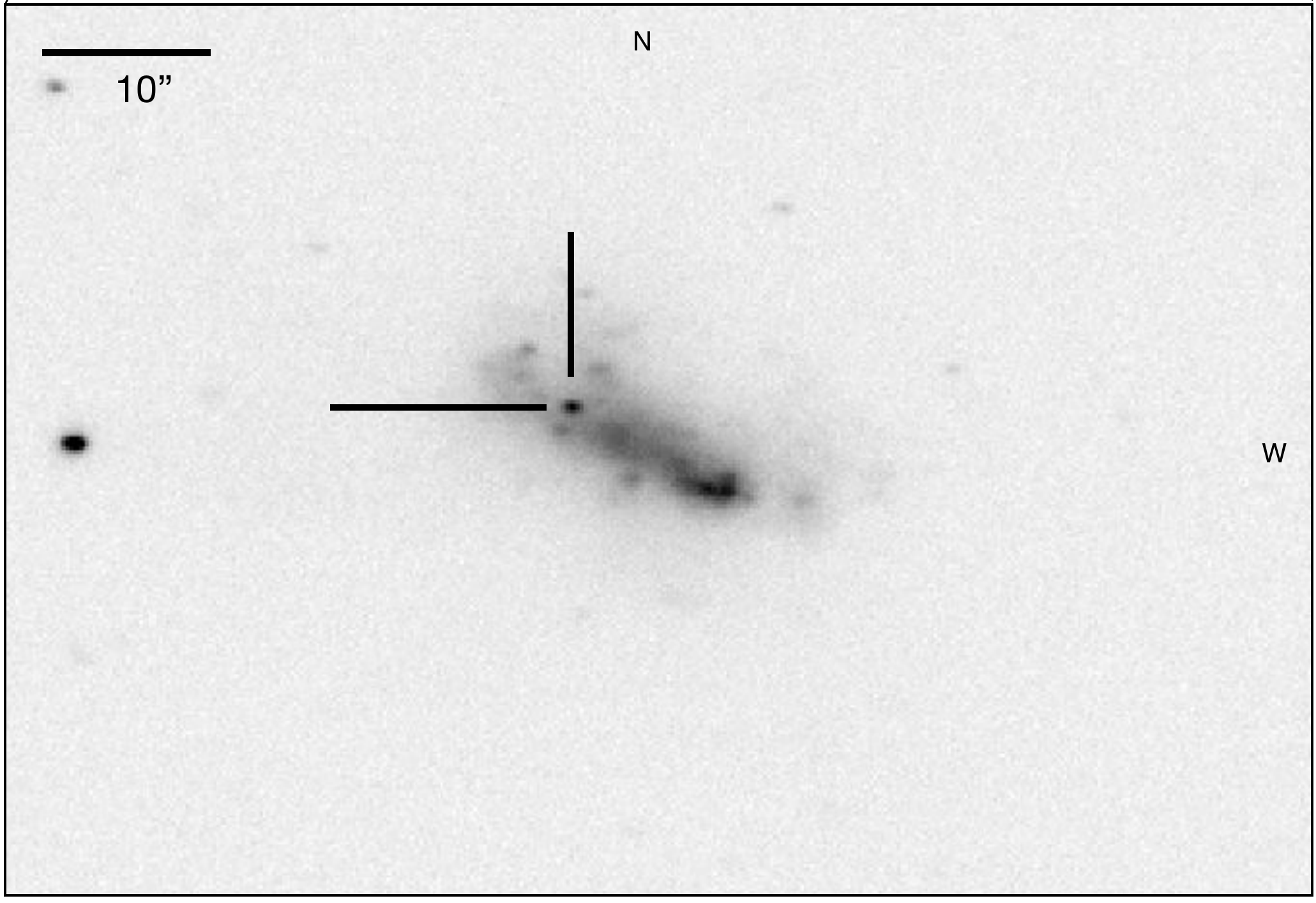}
    \caption{Finding chart of SBS 1436+529A (PGC052275) 
    constructed with a $g$-band image obtained with the NOT (+ ALFOSC)  on 2019 June 07. The position of SN~2019gsc is marked.}
    \label{fig:map}
\end{figure}

SN~2019gsc (also known as ATLAS19mbg, PS19bex and ZTF19aawhlcn) was discovered on 2019 June 02.35 UT by the ATLAS (Asteroid Terrestrial-impact Last Alert System) Project \citep{tonry2018}, at a cyan-ATLAS AB magnitude of 19.66 \citep[][see also Tonry et al. 2019, TNS Astronomical Transient Report No. 36575]{smartt2019}. The discovery report also noted a non-detection two days earlier (2019 May 31.39 UT) although in a different filter (orange-ATLAS) and not very deep (19.34 mag). The SN is located in the irregular galaxy SBS~1436$+$529A (other name: PGC052275), see Fig.~\ref{fig:map}, which is characterized by the presence of clumpy \ion{H}{ii} regions, similarly to SN~2008ha's host-galaxy UGG~12682. The galaxy is described as a merging system, with two nuclei separated by 3.8 arcsec, which corresponds to 0.9 kpc \citep{h2008}.

A heliocentric recessional velocity of $3384\pm3$ km~s$^{-1}$ for SBS~1436$+$529A is reported in the NASA/IPAC Extragalactic Database (NED) estimated from the Sloan Digital Sky Survey Data Release 3 \footnote{http://www.sdss.org/dr3/products/spectra/getspectra.html} observations. 
Adopting $H_{0} = 73.2 \pm 1.7$ km~s$^{-1}$ Mpc$^{-1}$ \citep{riess2016} and corrections for peculiar motions due to influences of the Virgo cluster, the Great Attractor, and the Shapley supercluster \citep{mould2000}, we obtain a Hubble flow distance and distance modulus to the host of SN~2019gsc of $53.4\pm 3.7$~Mpc and $\mu = 33.64 \pm 0.15$~mag, respectively. The NED database also provides a Milky Way reddening in the direction of SBS~1436$+$529A of $E(B-V)_{\textit{MW}} = 0.008 \pm 0.005$ mag \citep{schlafly2011}. We do not detect interstellar \ion{Na}{i}~D lines at the redshift of the host, therefore we assume that the reddening in the host of SN~2019gsc is minimal. 

A spectrum of SN~2019gsc was obtained a day after discovery, on 2019 June 03.97 UT \citep{leloudas2019} with the Nordic Optical Telescope 2.56m (NOT), and it was found very similar to that of SN~2010ae \citep{stritzinger14} at around the maximum light. SN~2019gsc exhibits a number of lines with low velocities, between 2,000$-$3,500 km~s$^{-1}$. In particular, the expansion velocity measured from the Si~II $\lambda$ 6355 line is $\approx$3,500 km~s$^{-1}$.

Given the relatively small number of well-observed SNe~Iax in the literature \citep{jha2017} and the close similarity of SN~2019gsc to the faintest members of this peculiar class of likely thermonuclear supernovae, we initiated a follow-up campaign to collect optical spectroscopic and photometric observations. Our follow-up began as soon as  SN~2019gsc  was classified. However, due to the low-luminosity and rapid evolution, the observing campaign was terminated just 54~d later (see Section~\ref{sec:late}). 

Basic information of SN~2019gsc and its host galaxy is summarized in Table~\ref{tab:infos}. The logbook of observations is reported in Tables~\ref{tab:photo} and \ref{tab:logbook}.

\begin{table}
	\centering
	\caption{Basic information on SN~2019gsc and its host-galaxy.}
	\label{tab:infos}
	\begin{tabular}{ll}
		\hline
Host galaxy &  SBS 1436+529A (PGC052275) \\
Galaxy type & Irregular \\
Heliocentric radial velocity & $3384\pm3$ km s$^{-1}$ \\
Redshift ($z$) & 0.011288 \\
Distance modulus & $33.64\pm0.15$ mag \\
Galactic extinction & $E(B-V)_{\textit{MW}} = 0.008 \pm 0.005$ mag \\
                    & ($A_{\text{B}}$ = 0.034 mag, $A_{\text{V}}$ = 0.026 mag) \\
                    \hline
SN spectral type & Iax \\
RA (J2000.0) & 14$^h$37$^m$45.24$^s$ \\
Dec (J2000.0) & +52$^{\circ}$43$'$36.43$''$  \\
Discovery date & 2019-06-02 08:31:12 UT \\ 
               & (ATLAS cyan mag 19.66) \\
Last no-detection & 2019-05-31 09:27:21 UT \\ 
Date of $g$-band maximum & 2019-06-01 (MJD 58635.4) \\
Date of $r$-band maximum & 2019-06-07 (MJD 58641.4) \\
$m_{g}$ at maximum & $19.92\pm0.01$ mag \\
$M_{g}$ at maximum & $-13.58\pm0.15$ mag\\
$m_{r}$ at maximum  & $19.46\pm0.01$ mag \\
$M_{r}$ at maximum & $-14.28\pm0.15$ mag\\
$\Delta m_{15}$(g) & $1.08\pm0.06$ mag \\
$L_{\text{bol}}$ at maximum & $\approx 10^{41}$ erg s$^{-1}$\\
\hline
	\end{tabular}
\end{table}

\subsection{Photometric data reduction}
\label{sec:data} 

Optical multi-band ($ugriz$) imaging of SN~2019gsc started soon after discovery and continued over the course of about fifty days, when the SN faded down to $m_r = 22.48 \pm 0.11$ mag. The telescopes and instruments used for the photometric follow-up campaign are listed in Table~\ref{tab:photo}.

All science images were pre-processed using standard procedures in \texttt{IRAF}\footnote{\texttt{IRAF} was distributed by the National Optical Astronomy Observatory, which is operated by the Association of Universities for Research in Astronomy (AURA) under cooperative agreement with the National Science Foundation.} for bias subtraction and flat fielding. The Sloan Digital Sky Survey \citep[SDSS; Data Release 15, ][]{aguado2019}\footnote{https://www.sdss.org/dr15} stars in the field of SBS~1436$+$529A were used as a local photometric reference. 

Accurate photometric measurements of SN~2019gsc require galaxy subtraction to isolate the SN flux contribution.
We therefore used the SDSS images obtained
on 2003 March 07 as templates to remove the host-galaxy
contamination from each of the $ugriz-$band science images. The final calibrated photometry (cf. Table~\ref{tab:photo}) was derived with the SNOoPY pipeline\footnote{SNOoPY is a package for SN photometry through PSF fitting and template subtraction developed by E. Cappellaro, 2014. A package description
can be found at http://sngroup.oapd.inaf.it/snoopy.html}, using the PSF-fitting technique
on the template-subtracted science images and the instrumental magnitudes were then calibrated to the Sloan AB system, using Sloan stars in the field of the host galaxy. 

\begin{figure}
    \centering
    \includegraphics[width=\columnwidth, angle=0, scale=0.8]{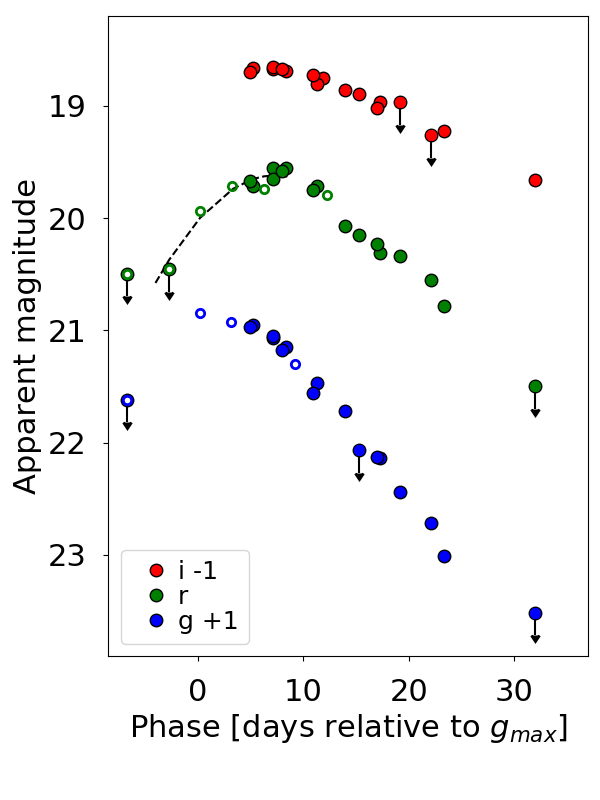}
    \caption{Optical $gri$-bands light curves of SN~2019gsc in the AB system. For clarity, the light curves have been shifted vertically as indicated in the legend. The open circles correspond to photometric points taken from the ZTF archive (cf. Table~\ref{tab:photo}). Epochs with arrows are upper-limits. The early portion of the $r$-band light curve is fitted with an {\it expanding fireball model}, with $n$ = 2 (cf. Section~3).}
    \label{fig:lc}
\end{figure}

Error estimates for the SN magnitudes are obtained through artificial star experiments in which fake stars with a similar magnitudes as the SN are placed in the fit residual image in a position close to, but not coincident with, the SN location. The simulated images are processed through the same procedure and the standard deviation of the measured magnitudes of the fake stars is taken as an estimate of the instrumental magnitude error. In practical term, this error mainly reflects uncertainty in the background subtraction.
The photometry is reported in Table~\ref{tab:photo} and plotted in Fig.~\ref{fig:lc}.\footnote{Note that since only five upper limits and two epochs are available for the $u$- and $z$-bands, respectively, their values are listed in Table~\ref{tab:photo} but not included in Fig.~\ref{fig:lc}.}

The early photometric epochs of SN~2019gsc obtained by the Zwicky Transient Facility survey \citep[ZTF]{kulkarni2018} are also added in our analysis. Data were retrieved from the Transient Name Server (TNS)\footnote{https://wis-tns.weizmann.ac.il/} and {\it Lasair}\footnote{https://lasair.roe.ac.uk/object/ZTF19aawhlcn/} \citep{smith2019}, listed in Table~\ref{tab:photo} and plotted in Fig.~\ref{fig:lc} as open circles. The cyan- and orange-ATLAS filters are non-standard; so the discovery and non-detection magnitudes mentioned at the beginning of Section~2 are not used in our analysis.

\begin{table*}
	\centering
	\caption{Optical photometry of SN~2019gsc. Sloan $ugriz$ filters (ABmag).}
	\label{tab:photo}
	\begin{tabular}{lcccccccc} 
		\hline
Date       & MJD & phase$^a$ &$u$ &  $g$ &$r$  & $i$& $z$ & Survey or \\
&     & & (mag)  & (mag)  & (mag) & (mag)& (mag) & telescope$^b$ \\ \hline
\multicolumn{3}{l}{\bf{Data from {\it Lasair}:}}&&&&\\
2019-05-25&58628.30&$-7$&$\cdots$ & $\cdots$ & $>$20.495 & $\cdots$ & $\cdots$ & ZTF\\
2019-05-25&58628.22&$-7$&$\cdots$&$>$20.617&$\cdots$&$\cdots$&$\cdots$& ZTF\\
2019-05-29&58632.28&$-3$&$\cdots$&$\cdots$&$>$20.455&$\cdots$&$\cdots$& ZTF\\
2019-06-01&58635.23&$0$&$\cdots$&19.848 (0.166) & $\cdots$&$\cdots$&$\cdots$& ZTF \\
2019-06-01&58635.26&$0$&$\cdots$&$\cdots$&19.933 (0.133) &$\cdots$&$\cdots$&ZTF \\
2019-06-04&58638.20&3&$\cdots$&19.922 (0.177) &$\cdots$&$\cdots$&$\cdots$& ZTF \\
2019-06-04&58638.26&3&$\cdots$&$\cdots$&19.717 (0.123)&$\cdots$&$\cdots$& ZTF \\
2019-06-07&58641.30&6&$\cdots$&$\cdots$&19.737 (0.111)&$\cdots$&$\cdots$& ZTF \\
2019-06-10&58644.20&9&$\cdots$&20.303 (0.287) &$\cdots$&$\cdots$&$\cdots$& ZTF \\
2019-06-13&58647.25&12&$\cdots$&$\cdots$&19.790 (0.171)&$\cdots$&$\cdots$& ZTF\\
\hline 
\multicolumn{2}{l}{\bf{Data from this work}:} &&&&&& \\
2019-06-05& 58639.98&5& $>$19.950 &19.970 (0.050) & 19.673 (0.054)& 19.701 (0.038)& $\cdots$ &    LT\\
2019-06-06& 58640.26&5&  $\cdots$&19.955 (0.032) & 19.716 (0.050)& 19.663 (0.071) &  $\cdots$  & LCO\\ 
2019-06-08& 58642.13&7& $\cdots$ &19.984 (0.029) & 19.650 (0.016)& 19.658 (0.021)& 19.940 (0.047)& NOT\\ 
2019-06-08& 58642.15&7& $\cdots$ &20.007 (0.049) & 19.551 (0.043)& 19.671 (0.067) &  $\cdots$  & LCO\\ 
2019-06-08& 58643.00&8& $>$20.292 & 20.175 (0.059)& 19.578 (0.043)& 19.673 (0.035)&$\cdots$&LT\\ 
2019-06-09& 58643.35&8&  $\cdots$&20.150 (0.400) & 19.553 (0.044)& 19.686 (0.073)&  $\cdots$ &  LCO\\ 
2019-06-11& 58645.96&11& $>$20.246 &20.561 (0.137) & 19.750 (0.061)& 19.725 (0.061)&  $\cdots$& LT\\ 
2019-06-12& 58646.32&11&  $\cdots$&20.472 (0.103) & 19.713 (0.076)& 19.807 (0.130)&  $\cdots$& LCO\\ 
2019-06-15& 58649.01&14& $>$19.283 &  20.715 (0.066)& 20.071 (0.131)& 19.863 (0.076)& $\cdots$& LT\\
2019-06-16& 58650.34&15& $\cdots$ & $>$21.068 & 20.151 (0.123)& 19.894 (0.139)&  $\cdots$&  LCO \\
2019-06-17& 58651.98&17& $>$19.247&  21.124 (0.107)& 20.231 (0.144)& 20.019 (0.037)&$\cdots$&LT\\
2019-06-18& 58652.24&17&  $\cdots$&21.140 (0.205) & 20.309 (0.173)& 19.964 (0.121) & $\cdots$& LCO\\ 
2019-06-20& 58654.21&19& $\cdots$ &21.438 (0.185) & 20.340 (0.147)& $>$19.967& $\cdots$& LCO\\ 
2019-06-23& 58657.15&22& $\cdots$ &21.718 (0.229) & 20.550 (0.208)& $>$20.264 &$\cdots$&LCO\\ 
2019-06-24& 58658.31&23& $\cdots$ &22.007 (0.188) & 20.779 (0.097)& 20.226 (0.080) & $\cdots$& LCO\\ 
2019-07-02& 58666.97&32& $\cdots$ & $>$22.513& $>$21.496 & 20.665 (0.029)& 20.175 (0.057) & NOT\\
2019-07-24& 58688.90&54& $\cdots$ & $\cdots$ & 22.480 (0.110) & $\cdots$ & $\cdots$ & GTC \\
\hline
	\end{tabular}
		\begin{tablenotes}
\item $^a$ Phase is relative to the epoch of  $g$-band maximum MJD = 58635.4.
\item $^b$ NOT = Nordic Optical Telescope, LT = Liverpool Telescope, LCO = Las Cumbres Observatory global network.
\end{tablenotes}
\end{table*}

\subsection{Spectroscopic data reduction}

\begin{table}
	\centering
	\caption{Log of spectroscopic observations of SN~2019gsc.}
	\label{tab:logbook}
	\begin{tabular}{lcccc} 
	\hline
Date & MJD &  phase$^a$ &   telescope$^b$ &  range (nm) \\ \hline
2019-06-03&58638.97&4 &NOT& $350\div920$\\
2019-06-11&58645.96&11 &NOT &  $350\div920$\\	
2019-07-24&58688.90&54&GTC& $350\div1000$ \\
\hline
	\end{tabular}
	\begin{tablenotes}
\item $^a$ Phase is relative to the epoch of $g$-band maximum MJD = 58635.4.
\item $^b$ NOT = Nordic Optical Telescope, GTC = Gran Telescopio Canarias.
\end{tablenotes}
\end{table}

Three low-resolution optical spectra of SN~2019gsc were obtained at phases +4~d, +11~d and +54~d relative to the epoch of $g$-band maximum (cf. Table~\ref{tab:logbook}). The early two spectra were obtained with the 2.56-m Nordic Optical Telescope (NOT): the first one under the program 59-509 (P.I. D. Malesani), made available in TNS; the second one via NUTS2 (NOT Un-biased Transient Survey 2) collaboration\footnote{http://csp2.lco.cl/not/} under the program 59-506 (P.I. E. Kankare et al.). The third late-time spectrum was collected with the 10.4-m Gran Telescopio Canarias (GTC), submitting a director discretionary program, DDT-GTC2019-129 (P.I. N. Elias-Rosa). 

We used standard procedures and \texttt{IRAF} tasks for the data reduction. After bias and flat-field correction, the SN spectrum was extracted and calibrated in wavelength with reference to standard arc lamp spectra. For the flux calibration, nightly sensitivity functions were derived from observations of spectrophotometric standard stars, used also to derive the corrections for the telluric absorption bands. A non-perfect removal of telluric absorption can affect the SN spectra, particularly at the position  of the strong O$_2$ A-band located between 7590$-$7650 \AA.  
Finally, the flux-calibration of each spectrum was verified against coeval broadband photometry.

\section{Light curve analysis}

The photometric evolution of SN~2019gsc in the better sampled $gri$ bands and for the first thirty days of evolution is plotted in Fig.~\ref{fig:lc}.
Only the $r$-band light curve shows the rise to maximum. The luminosity decline is slower at longer wavelength ($i$-band) than at shorter ones ($g$-band). By using a low-order polynomial fit to the optical $g$- and $r$-band light curves (the first $i$-band point is at maximum) we obtain the estimates of the magnitude and of the epoch of $g$ and $r$ maximum brightness, as well as the magnitudes reached 15~d after $g$ and $r$ peaks for the measure of the decline rate parameter $\Delta m_{15}$. Thanks to the inclusion of ZTF early-phase points, we find that SN~2019gsc reached an absolute magnitude at peak of $M_g = -13.58 \pm0.15$ mag ($m_g = 19.92 \pm 0.01$ mag) on MJD = 58635.4. The $r$-band peak occurred about $6$~d later at $M_r = -14.28 \pm0.15$ mag ($m_r = 19.46 \pm 0.01$ mag). 
The uncertainties in the absolute magnitudes are inferred propagating the errors of the fit to the peak apparent magnitudes and the errors in the adopted Galactic extinction and distance (cf. Section~2 and Table~\ref{tab:infos}). 
The decline rates of SN~2019gsc are: $\Delta m_{15}$($g$)~$= 1.08 \pm 0.06$ mag and  $\Delta m_{15}$($r$)~$= 0.96 \pm 0.06$ mag. 
In the case of normal SNe~Ia this parameter is known to correlate with luminosity \citep{phillips1993}, while for SNe Iax the relation exhibits significant scatter \citep{foley2013}. 
 Positioning SN~2019gsc on the decline rate \textit{versus} peak absolute $r$-band magnitude plot by \citet[][cf. their Fig. 2]{jha2017} or \citet[][cf. their Fig. 6]{magee2016}, we see that the fainter type Iax SNe~2019gsc, 2008ha and 2010ae are clustered on the bottom-right of the graph, evolving faster if compared to brighter SNe~Iax.  
 
 Light curve parameters for SNe~2008ha, 2010ae and 2019gsc are reported in Table~\ref{tab:parameters}. At peak brightness, SNe~2019gsc and 2008ha have comparable $r$-band magnitude (i.e. $-14.28\pm0.15$ mag \textit{vs} $-14.25\pm014$ mag), while SN 2019gsc seems to be slightly fainter in the $g$-band (i.e. $-13.58\pm0.15$ mag \textit{vs} $-13.89\pm0.14$ mag). These absolute magnitudes are calculated adopting $A_V = 0.026$ mag for SN~2019gsc (cf. Section~\ref{discovery}) and $A_V = 0.21$ mag for SN 2008ha (as in \citealt{stritzinger14} and in agreement with the Galactic extinction from the \citealt{schlafly2011} recalibration of infrared-based dust maps). 
 The estimate of the total reddening associated to SN~2010ae is uncertain. For a \textit{`moderate'} extinction $E(B-V)_{\text{ tot}} = 0.3$~mag which is used in this paper ($A_B = 1.2$ mag, $A_V =0.9$ mag; cf. \citealt{stritzinger14}; see also \citealt{lyman2018}), the $g$- and $r$-band absolute magnitudes for this supernova are $\sim -14.2$ and $-14.6$ mag, respectively.
 
\begin{figure}
    \centering
    \includegraphics[width=\columnwidth, angle=0, scale=1.0]{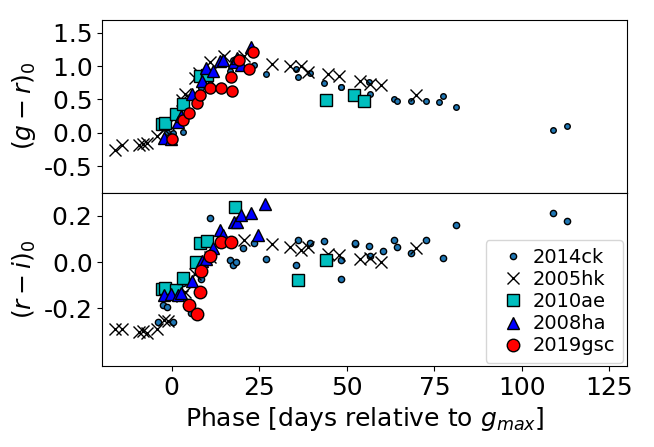}
    \caption{Extinction-corrected colour curves of SN~2019gsc compared with those of SNe 2008ha, 2010ae, and 2005hk \citep{stritzinger14,stritzinger15}, and SN~2014ck \citep{tomasella16}.}
    \label{fig:col}
\end{figure}

The extinction-corrected, $(g-r)_\textit{\/0}$ and $(r-i)_\textit{\/0}$ intrinsic colours of SN~2019gsc are plotted in Fig.~\ref{fig:col}. 
For comparison, the well-sampled, intrinsic colour curves of the type~Iax SNe~2008ha \citep{stritzinger14}, 2005hk \citep{phillips2007,stritzinger15}, 2010ae \citep{stritzinger14}, and 2014ck \citep{tomasella16} are also included. The colour-curves are corrected for the Galactic and host galaxy reddening
adopting extinction values of $A_V =$ 0.026 mag (SN~2019gsc), 0.21 mag (SN~2008ha), 0.35 mag (SN~2005hk), 0.9  mag (SN~2010ae), 1.5 mag (SN~2014ck), and a standard $R_V = 3.1$ reddening law \citep{cardelli1989}. 

We obtained an estimate of the rise time of SN~2019gsc by stretching and then matching the $r$-band light curve to that of SN~2005hk, which has densely sampled multi-colour light curves and a very well constrained rise time to maximum \citep[i.e. $\approx 15$ days in the $B$ band and $\approx 22$ days in the $r$ band; see][]{phillips2007}. Thus, using a time stretching factor of about 0.7, the $r$-band rise time of SN~2019gsc is estimated to be $\approx 15$ days.
Alternatively, a constraint of the rise time can be obtained by fitting the early-phase light curves of the SN, if pre-maximum epochs are available, as described by \cite{firth2015}. For SN~2019gsc the portion of the $r$-band light curve from pre-maximum to maximum (seven epochs, from MJD = 58635.26 to 58642.15 in Table~2) is fit with an \textit{expanding fireball} model $f_{model} (t) = \alpha (t-t_0)^n$, with $n = 2$ \citep{riess1999}, similarly to what was done for SN~2014ck \citep{tomasella16}. 
From our polynomial best-fit to the early light curve of SN~2019gsc, we obtain  $t_0 = 58626^{+3}_{-5}$ MJD as the time of the first light. The reported uncertainty corresponds to the standard deviation of the $t_0$ parameter when fitting a range of power laws, having $1.8\leq n \leq 2.2$, to the pre-maximum $r$-band light curve, as the analysis of large SN samples are found to be consistent with that spread of the $n$ index \citep{piro2014}. Consequently, SN 2019gsc has an $r$-band rise time $t = 16^{+3}_{-5}$ d, and a $g$-band rise time $\approx 10$ days ($10^{+3}_{-5}$~d), applying the $\sim$6 days lag between $g$ and $r$ maximum. This is consistent to the value inferred by using the 0.7 stretching factor. In the following, we adopt 10 days as the rise time of the bolometric light curve of SN 2019gsc (with $\sim$40\% uncertainty), considering our $g$-band rise and the $\Delta{t}_{rise}$ of bolometric \textit{vs.} filtered data \citep[cf.][their Table~2]{firth2015}. 

\subsection{Bolometric light curve and energetics}

\begin{figure}
	\includegraphics[width=\columnwidth, angle=0, scale=1.0]{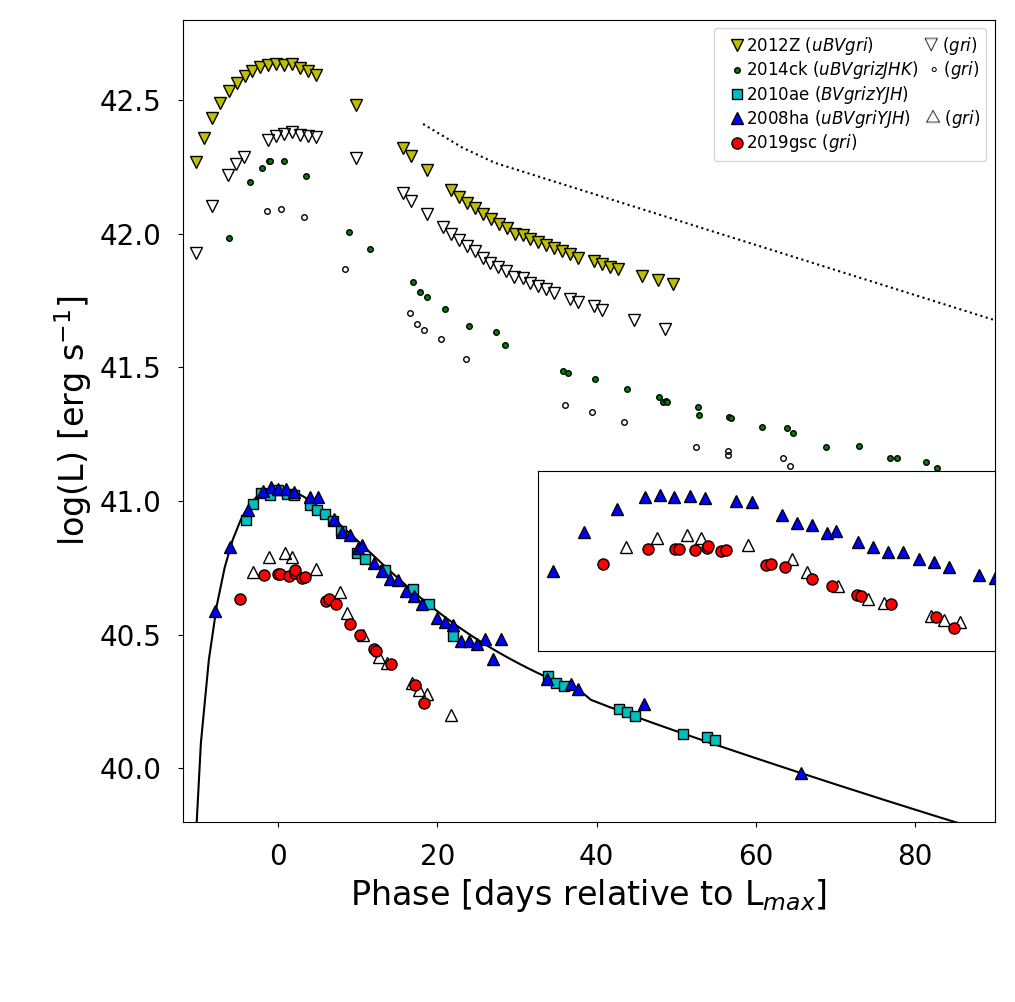}
    \caption{Pseudo-bolometric light curve of SN~2019gsc constructed by integrating over the $gri$-band flux points. The pseudo-bolometric light curves of SNe 2012Z (yellow triangles: $uBVgri$-bands; open triangles: $gri$-bands), 2014ck (green circles: $uBVgrizJHK$-bands; open circles: $gri$-bands), 2010ae ($BVgrizYJH$-bands), 2008ha (blue triangles: $uBVgriYJH$-bands; open trangles: $gri$-bands) are also plotted for comparison. The model fit of the SN 2008ha light curve is over-plotted as a solid line. The dotted line represents the SN~2012Z light curve with the inclusion of NIR photometry (see text for details). The inset shows a zoomed view around maximum of the pseudo-bolometric light curve of SN~2019gsc (red circles), compared with those of SN~2008ha constructed either including all the available $uBVgriYJH$-bands (blue triangles), or only the optical $gri$-bands (open triangles).}
    \label{fig:bol}
\end{figure}

Using the $gri$-band photometry of SN~2019gsc, we construct the pseudo-bolometric light curve in Fig.~\ref{fig:bol}.  
The photometry obtained at each epoch  was converted to flux at the effective wavelength of its corresponding passband. If photometry was not available in a given filter on a particular night, a magnitude was estimated through interpolation between adjacent epochs, or if necessary, by extrapolation assuming a constant colour from the closest available epochs. The fluxes were then corrected for reddening, yielding the spectral energy distribution (SED) at each epoch. These SEDs were integrated using a trapezoidal integration algorithm, assuming zero flux at the integration  boundaries. Finally, the flux at each epoch was converted to luminosities using our adopted distance to the host galaxy (see Table~\ref{tab:infos}). 

For comparison, the bolometric light curves of the faint SNe~2008ha and 2010ae, the intermediate SN~2014ck, and the bright SN~2012Z, are also plotted in Fig.~\ref{fig:bol}. The light curves of the comparison objects are computed following the same prescription adopted for SN~2019gsc, making use of the available photometry, reddening values (also reported above, cf. Section~3), and distances found in the literature, i.e.: ($i$) $E(B-V)_{\text{ tot}} = 0.11$ mag, $\mu = 32.59$ mag and $uBVgri$-bands for SN~2012Z (plus few $YJH$ late-epoch data), from \citealt{stritzinger15}; ($ii$) $E(B-V)_{\text{ tot}} = 0.5$ mag, $\mu = 31.94$ mag and $uBVgrizJHK$-bands for SN~2014ck, from \citealt{tomasella16}; ($iii$) $E(B-V)_{\text{tot}} = 0.07$~mag, $\mu = 31.55$ mag and $uBVgriYJH$-bands for SN~2008ha, from \citealt{stritzinger14}; ($iv$) $E(B-V)_{\text{ tot}} = 0.3$~mag, $\mu = 30.58$ mag and $BVgrizYJH$-bands for SN~2010ae, from \citealt{stritzinger14,lyman2018}; (with $H_{0} \equiv 73$ km~s$^{-1}$ Mpc$^{-1}$). Unfortunately, no ultraviolet (UV) nor near-IR (NIR) observations of SN~2019gsc were taken, and only a few $u$-band upper-limits were collected during our follow-up campaign. An estimation about the $uB$- and NIR-bands contribution to the light curve of SN~2019gsc is inferred from the ratio in flux between the Optical-to-InfraRed (OIR, i.e. using all the available $uBVgriYJH$-bands), and the optical ($gri$-bands only) bolometric light curves of SN~2008ha. The result is visually highlighted in the inset of Fig.~\ref{fig:bol}: the $gri$ light curves of SNe 2019gsc and 2008ha are almost overlapping, showing peak luminosity $L_{\text{19gsc,$gri$}} = 5.5 \pm 0.9 ~\times~10^{40}$ erg~s$^{-1}$ and $L_{\text{08ha,$gri$}} = 5.7 \pm 0.5 ~\times~10^{40}$ erg~s$^{-1}$, respectively. Thus we suppose that SN~2019gsc should have an OIR bolometric light curve very similar to the model fit of SN~2008ha, which is plotted as a solid line in Fig.~\ref{fig:bol} (peak luminosity $L_{\text{08ha,OIR}} \approx 1.12 \times 10^{41}$ erg~s$^{-1}$); i.e., applying a percentage increase by approximately 96\%, we derive $L_{\text{19gsc,OIR}} \approx 1.08 \pm 0.1 \times 10^{41}$ erg~s$^{-1}$.
The $u$-band alone contributes to $\sim$10\% at maximum (dropping to 1-2\% few days later); overall the bluest bands ($uBV$) account for most of the increase of the peak luminosity, while the NIR-bands ($YJH$) contribute $\sim10 \div 15\%$. This is roughly quantified comparing the fluxes of different pseudo-bolometric light curves of SN~2008ha, constructed including or excluding several photometric bands. The relevance of the NIR contribution to the late-phase bolometric light curve of SN 2012Z is also highlighted in Fig.~\ref{fig:bol}, where the OIR ($uBVgriYJH$) bolometric light curve is over-plotted as a dotted line to the optical ($uBVgri$) one (yellow triangles).
For what concern the ultraviolet (UV) contribution, we note that SNe~2005hk and 2012Z were detected in the UV at very early phases, but they both faded below the detection limit well before the epoch of optical maximum \citep{phillips2007,brown2009,stritzinger15}.
This implies that, in both objects, the flux in the UV drops well below 10\% of the total flux before maximum. In general, SNe~Iax are bluer than normal SNe~Ia in the UV before maximum light, but quickly redden (by $\sim$ $1.5-2$ mag in \textit{Swift uvw1-b}). Hence, $\sim$10 days post maximum, they are already redder than normal SNe~Ia \citep{jha2017}.

With the assumption that the light curve of SN~2019gsc is powered by $^{56}$Ni decay, the amount of $^{56}$Ni synthesized during the explosion is estimated using Arnett's rule \citep{arnett1982}. See \cite{stritzinger2005}, their Section 4, for an analytical expression that links $^{56}$Ni mass to the peak bolometric luminosity. Adopting 10$\pm$4 days rise time on the bolometric output of SN 2019gsc  
and $\approx 11\pm1~\times~10^{40}$ erg~s$^{-1}$ for the peak luminosity, with the  assumptions described in \citet{arnett1982}, we obtain the crude estimate of $M_{Ni} \approx  0.003 \pm 0.001 M_{\odot}$.
The relative error on $M_{Ni}$ is found by adding the relative errors on the peak luminosity (i.e. 0.10, which in turn depends on the accuracy of the adopted distance and reddening), and on the energy input from the decay of $^{56}Ni$ (i.e. 0.29; which is due to the error on the rise time).

Compared to normal SNe~Ia and to the prototype SN~2002cx, the low expansion velocity of the SN~2019gsc ejecta ($v_{\textit{ejc}} \approx 3.5 \times 10^{3}$ km s$^{-1}$, cf. Section~4 and Fig.~\ref{fig:temp_vel}) suggests it also has much lower ejecta mass ($M_{\textit{ej}}$) and kinetic energy ($E_{k}$). 
Quantitatively, the explosion parameters providing a good match to the bolometric light curve of SN~2019gsc are $M_{\textit{ej}} \approx 0.13 M_{\odot}$ and $E_{k} \approx 12 \times 10^{48}$ erg, using Arnett's equations \citealt{arnett1982} as done in \citealt{valenti2009} and \citealt{foley2009} for SN~2008ha. 
The energetics are similar to those inferred for both SN~2008ha and SN~2010ae. 
With regard to the reliability of the above estimations of ejecta mass and kinetic energy, we underline that the model relies upon a number of simplifying assumptions including: complete energy deposition supplied from
the radioactive $^{56}$Ni decay, no appreciable mixing of $^{56}$Ni, spherical symmetry and homologous expansion of the ejecta. Finally the model also relies on a constant optical opacity of about 0.1~cm$^2$ g$^{-1}$. It is well know that opacity has a strong dependence on the temperature and therefore  a strong time dependence \citep{hoeflich1992}. With these assumptions in mind, Arnett’s rule is accurate to within $\sim$50\% \citep{stritzinger15}.

\begin{table}
	\centering
	\caption{Light curves parameters for SN 2019gsc (this work), SN 2008ha, and SN 2010ae \citep[from][]{stritzinger14}. The absolute magnitudes for SNe 2019gsc,  2010ae and 2008ha are derived using $A_V= 0.026$, 0.90 and 0.21 mag, respectively.
	}
	\label{tab:parameters}
	\begin{tabular}{llll}
		\hline
Filter & $m_{peak}$ (mag) & $M_{peak}$ (mag) & $\Delta m_{15}$ (mag) \\
\hline
&&\bf{SN 2019gsc} & \\
$g$ & $19.92\pm0.01$ & $-13.58\pm0.15$ & $1.08\pm0.06$\\
$r$ & $19.46\pm0.01$ & $-14.28\pm0.15$ & $0.96\pm0.06$\\
\hline
&&\bf{SN 2010ae} & \\
$g$ & $17.49\pm0.02$ & $-14.2\pm0.5$ & $1.51\pm0.05$\\
$r$ & $16.92\pm0.02$ & $-14.6\pm0.5$ & $1.01\pm0.03$\\
\hline
&&\bf{SN 2008ha} & \\
$g$ & $17.97\pm0.02$ & $-13.89\pm0.14$ & $1.80\pm0.03$\\
$r$ & $17.57\pm0.01$ & $-14.25\pm0.14$ & $1.11\pm0.04$\\
\hline
	\end{tabular}
\end{table}

\section{Spectroscopic analysis}

\begin{figure}
	\includegraphics[width=\columnwidth, angle=270, scale=0.81]{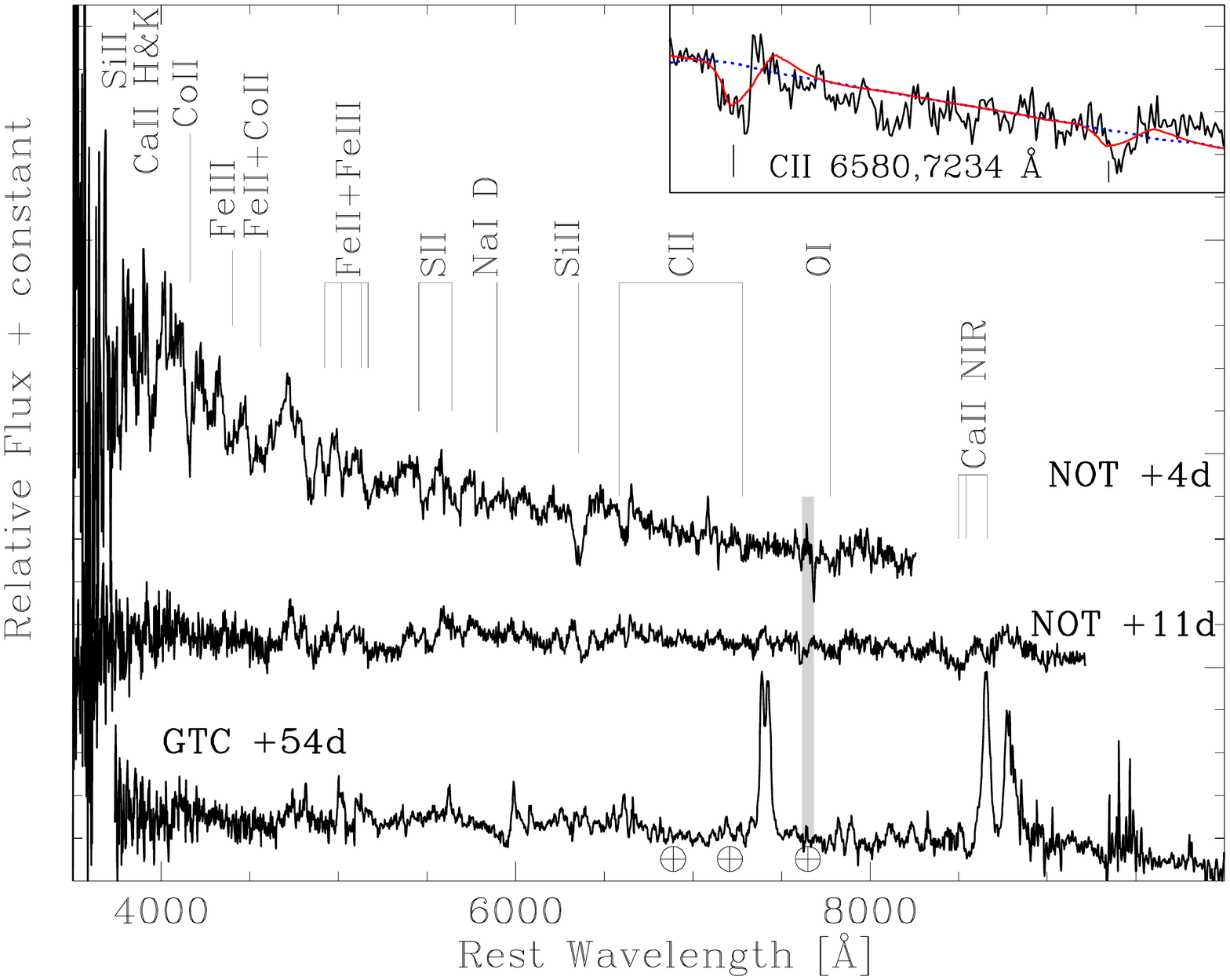}
    \caption{Spectroscopic sequence of SN 2019gsc, along with line identification of prominent features. The inset on the top shows the region of the +4d spectrum (black) centered on \ion{C}{ii} $\lambda\lambda$6580, 7234, with the \texttt{SYNOW} synthetic spectra of the \ion{C}{ii} (solid red line) and \ion{Fe}{ii} ions (dotted blue) overplotted.}
    \label{fig:spectral_seq}
\end{figure}

Our spectroscopic time series of SN~2019gsc consists of three optical spectra, distributed over fifty days. 
The last spectrum was taken when the object had faded to a magnitude of $m_r = 22.48 \pm 0.11$ mag. The spectroscopic sequence is shown in Fig.~\ref{fig:spectral_seq}. 

The first spectrum taken at maximum light exhibits a blue continuum (the blackbody temperature is $T_{\textit{bb}} \approx 9 \times 10^{3}\,$~K, cf. Fig.~\ref{fig:temp_vel}) superposed with a rich structure, along with several low-velocity P~Cygni features associated with both intermediate-mass elements (IMEs) and Fe-group elements. By comparison with normal SNe~Ia and using the \texttt{SYNOW} \citep{branch2004} spectral synthesis code for fitting our spectra with profiles of various ions, we identify: a relatively weak and narrow \ion{Si}{ii} $\lambda$6355 feature, along with absorptions attributed to \ion{Ca}{ii} H\&K and \ion{Ca}{ii} NIR triplet;  \ion{S}{ii} $\lambda\lambda$5454, 5640; \ion{Na}{i}~D; \ion{O}{i} $\lambda$7774 (which is visible from day +1 to +52); high-excitation lines of \ion{Fe}{ii}, \ion{Co}{ii} and \ion{Fe}{iii}, which appear to produce most of the observed features blueward of $\sim$5000 \AA\/; and, possibly, \ion{C}{ii} $\lambda\lambda$6580, 7234. There is some indication that early spectra of type Iax SNe contain \ion{C}{ii} and, sometimes, also \ion{C}{iii} \citep[i.e.][]{chornock2006,foley2013,tomasella16}. Actually, we obtain a reasonable fit to absorption features at $\sim 6580$ and 7230 \AA\/ adding the \ion{C}{ii} ion in the \texttt{SYNOW} spectral model (rather than other ions, i.e. \ion{Fe}{ii}), as it is shown in the inset of Fig.~\ref{fig:spectral_seq}.
If this identification is correct, the detection of \ion{C}{ii} in the ejecta of SN~2019gsc might be related to unburnt material of the progenitor system, giving indication on its nature \citep[C/O WD {\it vs} O/Ne/Mg WD; see][]{nomoto2013}, or on the mechanism by which the explosive flame propagates throughout the WD \citep{folatelli2012}. 

\begin{figure}
	\includegraphics[width=\columnwidth, angle=0, scale=1.0]{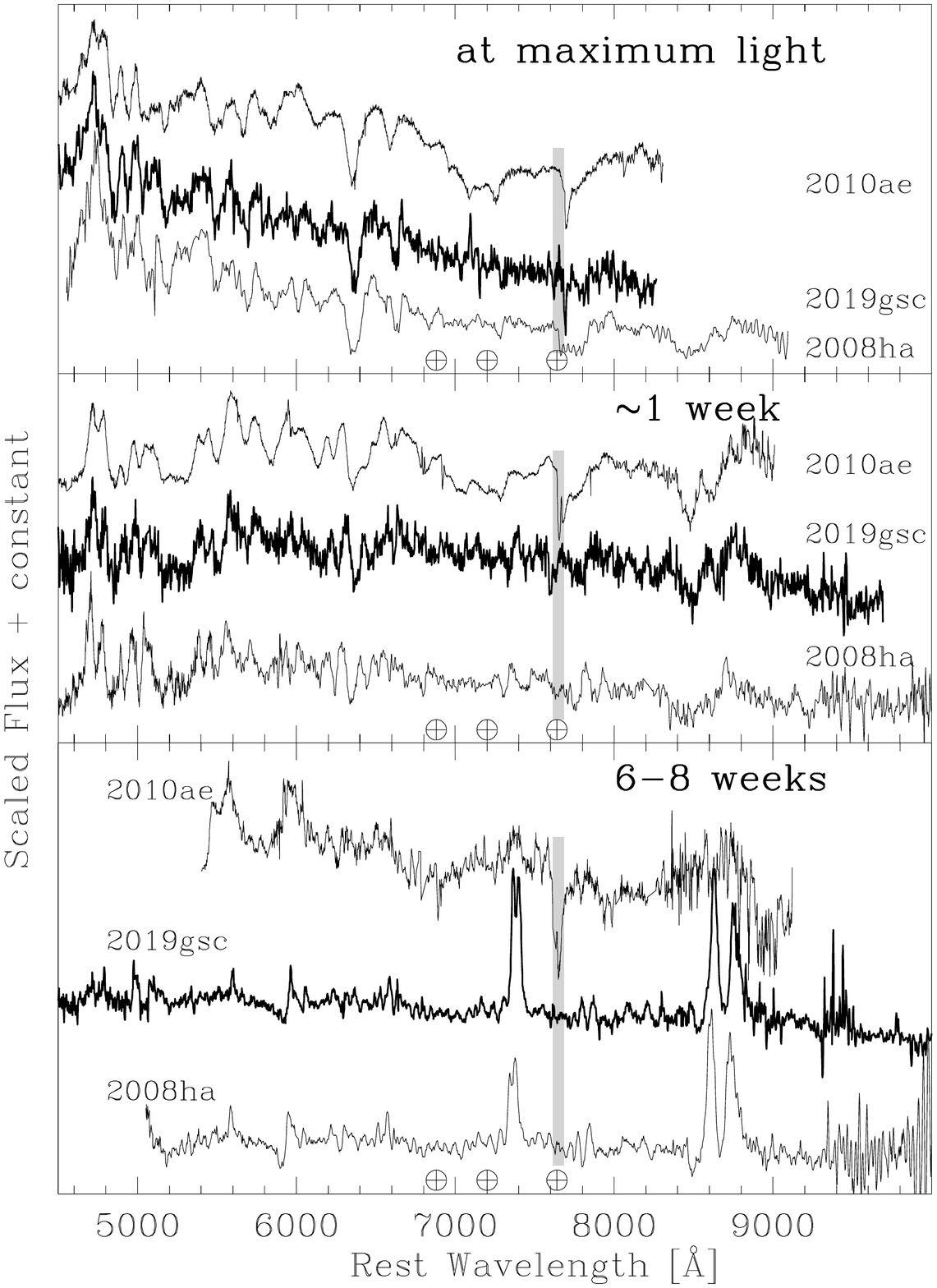}
    \caption{Sequences of spectra of the faint type~Iax SNe 2019gsc, 2010ae, and 2008ha at similar phases, extending from around maximum light, to one week, and finally out to $6 \div 8$ weeks post maximum. The spectra of SN~2008ha are from \citet{valenti2009} and those of  SN~2010ae from \citet{stritzinger14}.}
    \label{fig:spectral_match}
\end{figure}

In Fig.~\ref{fig:spectral_match} the spectra of SN~2019gsc are compared with those of the fast-and-faint type~Iax SNe 2008ha and 2010ae at similar phases. Notably, all spectra are very similar at maximum light and one week later, while about fifty days after maximum, forbidden calcium  emission lines are strong in SNe 2019gsc and 2008ha, while  in SN 2010ae they are relatively weak and emerge only during later epochs (see Section 4.1).

\subsubsection{Search for helium in +11 day spectrum}

\begin{figure}
	\includegraphics[width=\columnwidth, angle=270, scale=0.72]{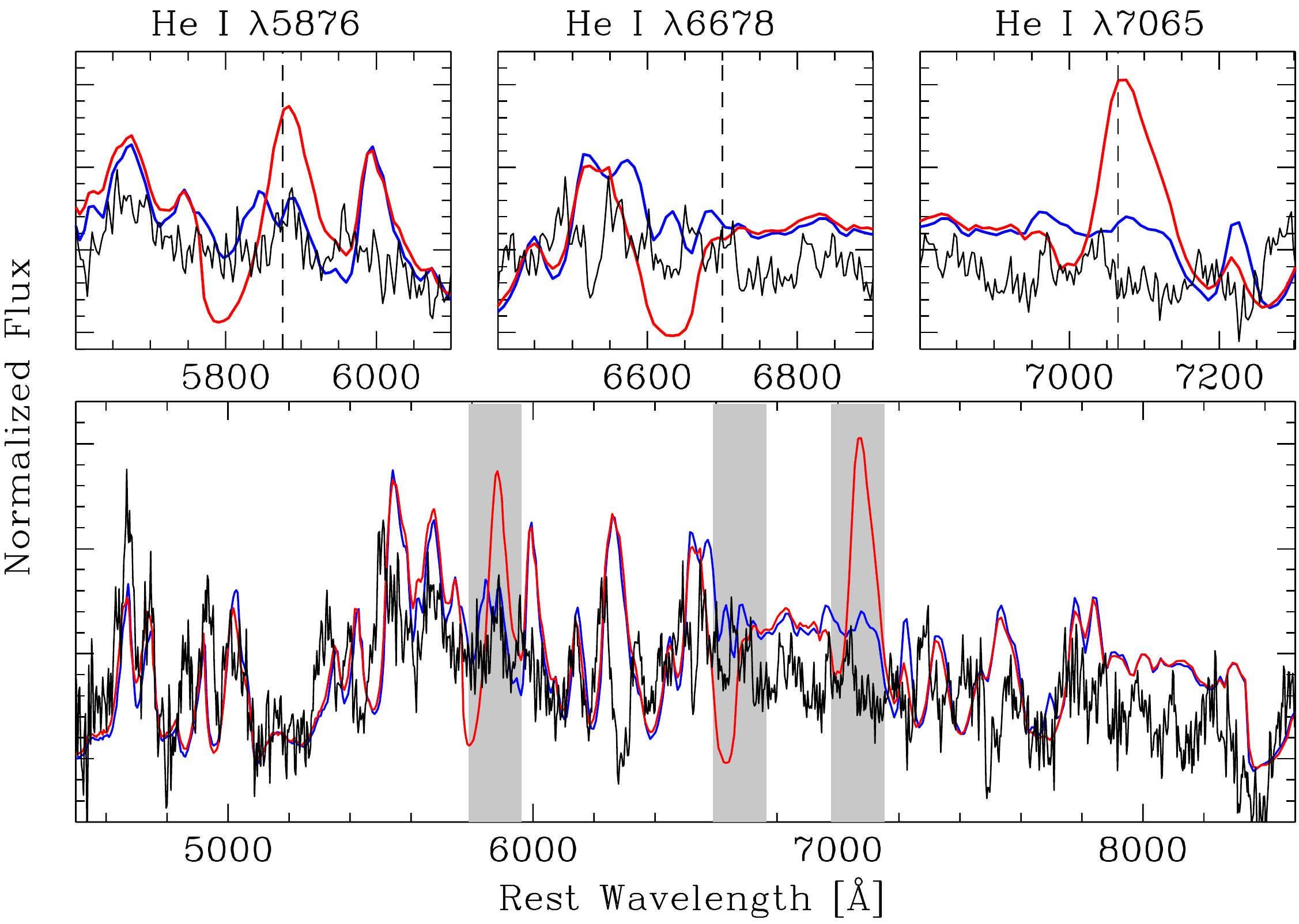}
    \caption{Comparison of the $+$11~d spectrum of SN 2019gsc with the N5def-hybrid models by \citet{magee2019} with 0.22\% (blue line) and 36\% (red line) helium abundances. On top, from left to the right the panels show the wavelength regions of \ion{He}{i} $\lambda$5876, $\lambda$6678 and $\lambda$7065, respectively (highlighted with shaded grey bars on the bottom panel). Spectra have been normalised to the median flux in the wavelength range 5000~\AA\/ $-$ 7000~\AA\/}.
    \label{fig:he_model}
\end{figure}

Several authors \citep[i.e.][]{filippenko2007,foley2013,foley2016,magee2019,jacobson2019} have investigated the prospect to detect helium features in SNe~Iax spectra, following the notion that their progenitor are C/O (or hybrid C/O/Ne) WDs accreting material from a helium star companion. \cite{foley2013} identified two SNe Iax with clear \ion{He}{I} lines in their post-maximum spectra, i.e. SN~2004cs and SN~2007J (cf. their Section 5.2 and Figure~22). They roughly estimate that $\sim$15\% of SNe Iax can exhibit helium signatures during a specific phase of their spectroscopic evolution. \cite{magee2019} presented a series of simulations with varying helium
abundances, calculated using the \texttt{TARDIS} radiative transfer code \citep{kerze2014} and based on the multi-dimensional explosion models of \cite{kromer2015}. They show that the best chance for detecting \ion{He}{i} features is either at NIR wavelengths (with \ion{He}{i} $\lambda$10830 being the strongest line) or in the post-maximum spectra of the faintest members of the class. Therefore, the extremely faint SN~2019gsc offers an excellent opportunity to search for signatures of helium, although our spectral coverage includes only the optical helium lines (i.e. \ion{He}{i} $\lambda$5876, $\lambda$6678, $\lambda$7065).

In Fig.~\ref{fig:he_model} we compare our +11~d optical spectrum of SN 2019gsc with the synthetic N5def-hybrid models calculated by \cite{magee2019} specifically for the faintest members of the SN~Iax class at +15~d after bolometric maximum light. As in \cite{magee2019}, spectra have been normalised to the median flux in the wavelength range 5000~\AA\/ $-$ 7000~\AA\/.
Only a marginal evidence of helium, if any, can be advanced for SN 2019gsc. The best match is with the N5def-hybrid model (blue line in Fig.~\ref{fig:he_model}) which has a low helium abundance (0.22\% of the ejecta mass). We note that the N5def-hybrid ejecta mass \citep[0.014 M$_{\odot}$;][]{kromer2015} is by an order of magnitude lower than the one estimated for SN~2019gsc (0.13 M$_{\odot}$; see Section~3.1). Hence, we speculate that SN~2019gsc could contain a few 10$^{-4}$ M$_{\odot}$ of helium at most.

\subsubsection{Expansion velocities of the ejecta}

\begin{figure}
	\includegraphics[width=\columnwidth, angle=0, scale=1.0]{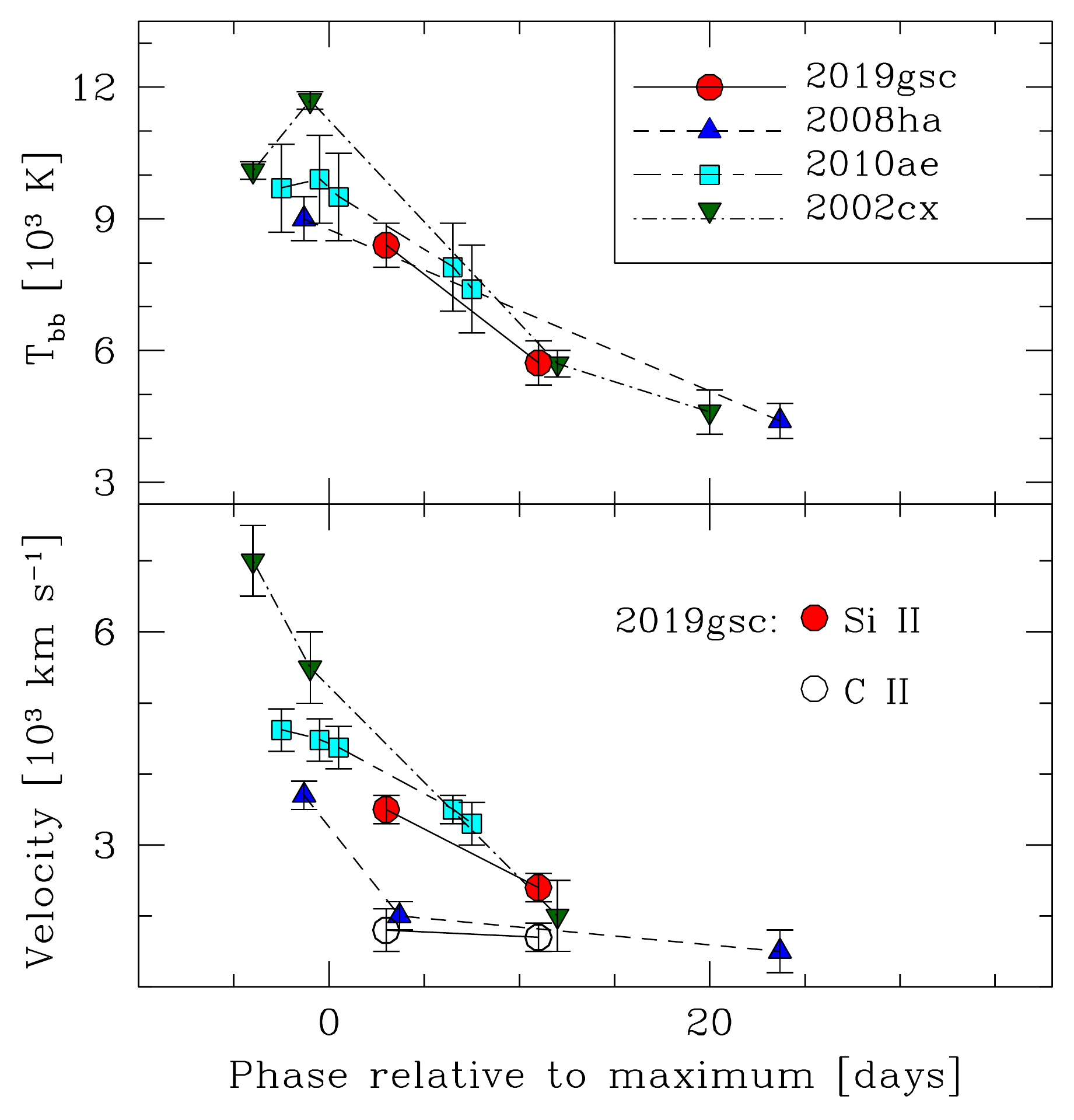}
    \caption{Top-panel: comparison of the time evolution of $T_{bb}$ for SN 2019gsc with SNe 2008ha, 2010ae and 2002cx. Bottom-panel: expansion velocity evolution of the ejecta of SN 2019gsc, estimated measuring the location of the absorption minimum of \ion{Si}{ii} $\lambda$6355 (filled red circles) and \ion{C}{II} $\lambda$6580 (open circles). \ion{Si}{ii} $\lambda$6355 velocity for SN~2008ha (filled blue triangles), SN~2010ae (filled cyan squares) and SN~2002cx (filled green triangles) are also reported.}
    \label{fig:temp_vel}
\end{figure}

The low expansion velocity of the ejected material is a hallmark of SNe~Iax and is indicative of a low kinetic energy in comparison to normal SNe~Ia explosions, as previously noted in Sect.~3.1. The ejecta velocity of SN~2019gsc is estimated from the location of the absorption minimum of \ion{Si}{ii} $\lambda$6355, which suffers negligible line blending relative to other features in the spectrum.
We measure a \ion{Si}{ii} expansion velocity of $v_{\textit{ejc}} \approx$ 3,500~km~s$^{-1}$ at the time of $g$-band maximum and this decreases to $v_{\textit{ejc}} \approx$~2,400~km s$^{-1}$ about ten days later. In Fig.~\ref{fig:temp_vel} (bottom panel) the expansion velocity evolution of the three faintest SNe~Iax, SNe~2008ha, 2010ae, and 2019gsc, is compared with the prototypical type~Iax SN~2002cx. We conclude that the expansion velocities of SN~2019gsc are similar to those of SN 2008ha, and about $\sim20$\% and $\sim40$\% lower than in SN~2010ae and SN~2002cx, respectively. 

Taking the \ion{Si}{ii} $\lambda$6355 absorption line as an indicator of the expansion velocity, the ratio between the Doppler velocity of the putative \ion{C}{ii} $\lambda$6580 and \ion{Si}{ii} $\lambda$6355 at maximum is $\sim$ 0.5 (cf. Fig.~\ref{fig:temp_vel}), which is similar to that observed in SN~2008ha \citep{parrent2011}. This ratio is generally slightly above unity in normal SNe~Ia \citep{folatelli2012}; $\sim$0.89 and $\sim$0.95 in SN~2012Z and SN~2014ck, respectively \citep{stritzinger15,tomasella16}. The fact that the \ion{C}{ii} Doppler velocity is significantly lower than the expansion velocity may indicate ejecta asymmetries as suggested by \citet{foley2016} for SN~2008ha, or alternatively the line is not associated with \ion{C}{ii}. In passing, we note that the \ion{C}{ii} line velocity is considerably lower than \ion{Si}{ii} velocity also for the peculiar 09dc-like SNe \citep[][see their Fig.~8]{taub2017,taub2019}.

Also plotted in Fig.~\ref{fig:temp_vel} (top panel) are the rough estimates of the photospheric temperature of SN 2019gsc as derived from a blackbody (BB) function fit to the spectral continuum. Only the first two epochs were considered, as at +54~d the presence of emission lines and line blanketing drive a flux deficit at shorter wavelengths. Spectra were corrected for redshift and extinction before the fit. Errors are estimated from the dispersion of measurements obtained with different choices for the spectral fitting regions. The photospheric temperature of SN~2019gsc at maximum is around 8,000 K to 9,000~K, rapidly decreasing below 6,000~K one week later, similar to the BB temperature evolution of SNe~2008ha and 2010ae.

\subsection{The Gran Telescopio Canarias late-time spectrum}
\label{sec:late}


\begin{figure*}
\includegraphics[width=\columnwidth, angle=0,scale=1.8]{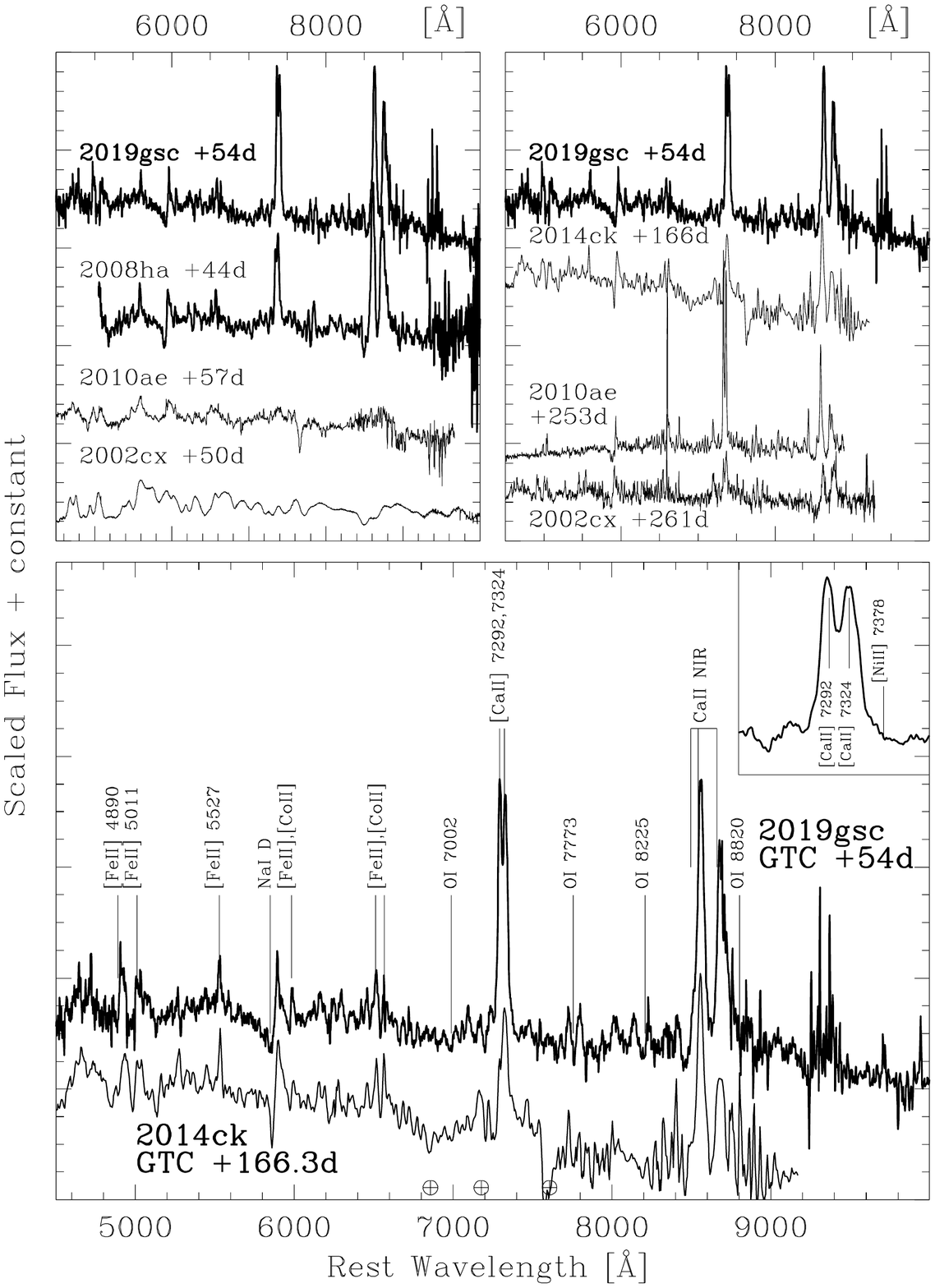}
\caption{Top-left panel: comparison of the $+$54~d GTC spectrum of SN~2019gsc with similar phase spectra of SN~2008ha \citep{valenti2009}, SN~2010ae \citep{stritzinger14}, and the prototype of the class SN~2002cx \citep{li2003}. Top-right panel: the same spectrum of SN~2019gsc is compared with the spectra of the brighter type~Iax SN 2014ck \citep{tomasella16} and SN~2002cx, and with the faint SN~2010ae, taken at phases $>$ 5 months. Bottom-panel: zoomed-view of the GTC late-spectrum of SN~2019gsc, with tentative identification of the principal forbidden lines, based on \citealt{li2003}. Permitted \ion{Ca}{II} 
and \ion{O}{I} lines (showing P~Cygni profiles) are also reported, while several \ion{Fe}{II} features (which are too numerous to be included) are superimposed on a pseudo-continuum, according to late-time spectral modelling of type Iax SNe~2002cx and 2005hk by \citealt{jha2006} and \citealt{sahu2008}, respectively. The inset shows the region centered on [\ion{Ca}{ii}] $\lambda\lambda$7292, 7324 doublet. The position of (weak/absent) [\ion{Ni}{ii}] $\lambda$ 7378 is marked.}
\label{fig:three_panels}
\end{figure*}

We observed SN 2019gsc with the Gran Telescopio Canarias (GTC) equipped with OSIRIS and using the R1000B$+$R1000R grisms. 
This setup provided wavelength coverage from 363.0 nm to 1000 nm. The  observations were carried out on 2019 July 24, when the SN was +54~d post-maximum and its brightness had faded to $m_r = 22.48\pm0.11$ mag (as measured in the acquisition images), with a total integration time of 2 hours. 

The high-quality GTC spectrum of SN~2019gsc exhibits the late-time spectral characteristic of the SN~Iax class, consisting of a weak continuum, or pseudo-continuum, superposed with both permitted and forbidden emission lines \citep{foley2016}. The best match is with the faint type Iax SN~2008ha at a similar phase (cf. Fig.~\ref{fig:three_panels}, top-left panel).

As shown in Fig.~\ref{fig:three_panels}, the late-time spectrum of SN~2019gsc is also similar to that of the brighter type~Iax SN~2014ck (peak $M_V \approx -17.29$ mag, \citealt{tomasella16}) taken at a phase  of +166.3~d, thus highlighting the very rapid spectroscopic evolution of the fainter members of the SN~Iax class.

Following \citet{li2003,jha2006,sahu2008}, we tentatively identify narrow, permitted \ion{Fe}{ii} features and a number of forbidden lines associated to iron, cobalt and calcium (see Fig.~\ref{fig:three_panels}). Several lines blueward of $\sim$6000 \AA\/ can be a blend of permitted \ion{Fe}{ii} features, which are identified at earlier epochs with higher velocities. Iron lines can be present also between 6200 and 6600 \AA\/, as highlighted by \cite{jha2006} and \cite{sahu2008} for type Iax SNe~2002cx and 2005hk, respectively.

The strongest features are attributed to calcium, including both the \ion{Ca}{ii} NIR triplet and the forbidden [\ion{Ca}{ii}] $\lambda\lambda$7292, 7324 doublet, having a full-width-half-maximum (FWHM) velocity  of ~1400 and 1500 km~s$^{-1}$, respectively. Both permitted and forbidden calcium lines are significantly more prominent in SN~2019gsc than SN~2002cx taken at a similar phase, and are comparable to those of SN~2008ha.

As for SNe 2002cx \citep{jha2006} and 2005hk \citep{sahu2008}, several weak \ion{O}{i} lines are tentatively recognized in SN~2019gsc until late phases (see Fig.~\ref{fig:three_panels}, bottom-panel), suggesting that the elements in the ejecta are homogeneously mixed. 
In contrast, forbidden oxygen features [\ion{O}{i}] $\lambda\lambda$6300, 6364 are not detected in this late-time spectrum of SN 2019gsc (so far [\ion{O}{i}] has never been observed in SNe Iax, cf. \citealt{jha2006,mccully2014,foley2016}). It was suggested that [\ion{Ca}{ii}] $\lambda\lambda$7292, 7324 can limit the strength of [\ion{O}{i}] $\lambda\lambda$6300, 6364 emission from a region in which both these ions co-exist \citep[][and references therein]{dessart2015}.
In passing, we note that normal stripped-envelope, core-collapse SNe exhibit strong [\ion{O}{i}] emission at late times. The absence of forbidden oxygen in the late-time spectra of SNe~Iax weakens the link with core-collapse SNe which have been also considered a possible explanation for SNe~Iax \citep{valenti2009,moriya2010}. 

To summarise, we find that the $+54$~d spectrum of SN~2019gsc is similar to those of other SNe~Iax at least half a year post maximum. This is an exceptionally fast spectral evolution, similar to  SN~2008ha, but not to  SN~2010ae \citep{stritzinger14}, which does not exhibit forbidden \ion{Ca}{ii} emission at the same phase (cf. Fig.~\ref{fig:three_panels}). 

\section{Environments of faint Type~Iax SNe}

SNe~Iax are usually found in star-forming, late-type host galaxies. \cite{lyman2018} have recently performed spectroscopic environmental measurements  of the host galaxies and the explosion sites for about twenty SNe~Iax (see their Table 1). They find that the metallicity distribution of the SNe Iax explosion site is similar to that of core-collapse SNe, and metal poor in comparison to normal SNe~Ia as well as spectroscopically classified 1999T-like objects. Moreover, fainter members of the SN~Iax class are found to be located in very metal poor environments. However, their sample contains a small number of very faint objects 
and consequently the statistics are limited. New SNe~Iax are crucial to gain further information on this topic, and, in this respect, the extremely-faint SN~2019gsc is particular interesting. 

In order to derive the metallicity  for the host galaxy of SN~2019gsc and for its explosion site, we use spectra downloaded from SDSS DR12 as well as our own GTC spectrum (cf. Sect.~4.1), and the calibration of \cite{pettini2004} or \cite{marino2013}. These authors calibrate the oxygen abundances of extra-galactic \ion{H}{ii} regions for which the oxygen abundance was determined with the direct $T_e$-based method against the {\it N2} and {\it O3N2} indices: $N2 \equiv$ log$_{\rm 10}$([\ion{N}{ii}]$\lambda$6583/H$\alpha$) and $O3N2 \equiv$ log$_{\rm 10}$([\ion{O}{iii}]$\lambda$5007/H$\beta$)/([\ion{N}{ii}]$\lambda$6583/H$\alpha$). The {\it N2} and {\it O3N2} indices have advantage over other types of calibrations in the literature due to the small wavelength separation between the lines, and hence are less sensitive to the extinction and to issues related to flux calibration. 

When using the archive spectra of SBS~$1436+529$A, the emission line flux measurement \citep[as described in][their Section 3.3]{lyman2018}, provides the indices {\it N2} = $-$1.43$\pm$0.03 and {\it O3N2} = 2.04$\pm$0.04. Thus, we estimate the SN~2019gsc host-galaxy oxygen abundance to be: $(i)$ 12 + log$_{\rm 10}$(O/H) $= 8.08 \pm 0.01$ dex, using the calibration for $O3N2$ of \cite{pettini2004};  or $(ii)$ 12 + log$_{\rm 10}$(O/H) $= 8.10 \pm 0.01$ dex, using the calibration for $O3N2$ of \cite{marino2013}. The reported errors are formal errors. Added in quadrature with systematic errors related to determining the gas-phase metallicity, i.e., $\approx 0.16-0.18$ dex \citep{marino2013}, we obtain 12 + log$_{\rm 10}$(O/H) $= 8.10 \pm 0.18$ dex (using the {\it O3N2} calibration by \citealt{marino2013}; similar values, within the error, come from the {\it N2} index and/or \citealt{pettini2004} calibration). This value is sub-solar \citep[using 12 + log$_{\rm 10}$(O/H) $= 8.69$ dex as the solar abundances, see][]{asplund2009}. 

\begin{figure}
     \centering
	\includegraphics[width=\columnwidth, angle=0, scale=1.0]{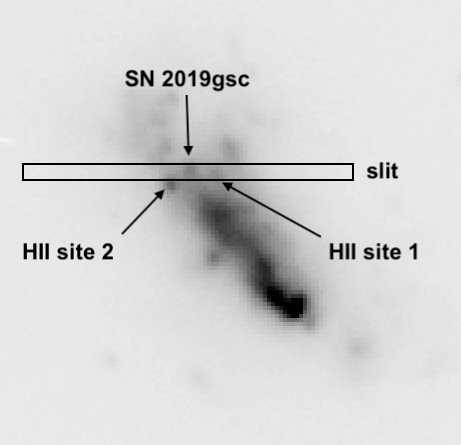}
    \caption{Slit position during the acquisition of the GTC spectrum of SN~2019gsc. Bright \ion{H}{ii} regions on the right (site 1) and on the left (site 2) of the SN were also captured.  }
    \label{fig:gtc_slit}
\end{figure}

As our GTC long-exposure (2 hours), long-slit spectrum includes also the bright \ion{H}{ii} regions both on the left and on the right of the position of the SN location (see Fig.~\ref{fig:gtc_slit}), we extracted a 1-D spectrum for each \ion{H}{ii} region. Thus, we derive also the oxygen abundance for the \ion{H}{ii} regions close to the SN~2019gsc explosion site using the individual line fluxes, in the same manner as done with the host-galaxy spectrum. 
We report the results in Tab.~\ref{tab:metal} and note that they too suggest a sub-solar `metallicity'. All three of our oxygen abundance estimates are in agreement within the errors.

  \begin{table}
	\centering
	\caption{Estimates for SN 2019gsc host-galaxy and explosion-site oxygen abundance 12 + log$_{\rm 10}$(O/H), using the calibration for $O3N2$ by Marino et al. (2013).}
	\label{tab:metal}
	\begin{tabular}{ll}
		\hline
host galaxy nucleus&  $8.10 \pm 0.18$ dex\\
SN site 1 & $8.23\pm 0.18$ dex\\
SN site 2& $8.17\pm 0.18$ dex\\
        \hline
	\end{tabular}
\end{table}

These measurements place SN~2019gsc on the metal-poor tail of the oxygen abundances distribution for the SN~Iax collection by \citet[][cf. their Table~4]{lyman2018}. The metallicity of the host galaxy of SN~2019gsc is lower than those of all SN Iax host galaxies in the \citet{lyman2018} sample, both for the SN explosion site and the host nucleus. It is very similar to the metallicity deduced from the explosion site of SN~2008ha, 
i.e. $8.22 \pm 0.18$ dex.

As a test case, we also downloaded the host spectrum of SN~2010ae, ESO 162-017 (from 6DF Galaxy Survey, via NED\footnote{http://www.6dfgs.net/}), and the estimation of the oxygen abundance is $8.32 \pm 0.18$ dex, using the $O3N2$ index and \cite{marino2013} calibration. This is in full agreement with the value of $8.34 \pm 0.14$ dex measured by \cite{stritzinger14} and $8.31 \pm 0.18$ dex by \cite{lyman2018}.

\section{Conclusions}     

The observations presented in this paper 
provide a view on an extreme member of the class of faint-and-fast evolving SNe~Iax that includes SN 2008ha, SN 2010ae, and even SN~2009J \citep{foley2013}.
SN~2019gsc is characterized by an extremely low luminosity, reaching an absolute peak $g$-band magnitude of only $M_g = -13.58 \pm0.15$ mag ($M_g = -13.89 \pm0.14$ mag for SN~2008ha), displays a decline rate of about $\Delta m_{15}$($g$)~$= 1.08\pm0.06$, and a $g$-band rise time of $\approx 10^{+3}_{-5}$~d. It also exhibits a low expansion velocity of the ejecta, about $\sim$ 3,000 km s$^{-1}$ at maximum, and consequently low kinetic energy, similar to SN~2008ha. The peaks of the pseudo-bolometric light curves indicate that SNe~2008ha, 2010ae and 2019gsc produced around a few $\times 10^{-3}$ M$_{\odot}$ of $^{56}$Ni. 
We report the simultaneous publication of observational data on SN 2019gsc by \cite{srivastav2020}. They find peak absolute magnitudes, decline rate (i.e. $M_g = -13.75 \pm0.23$ mag; $M_r = -13.97 \pm0.16$ mag; $\Delta m_{15}$($r$)~$= 0.91\pm0.1$ mag), and explosion parameters 
which are compatible, within the errors, with ours.

In many ways, SN~2019gsc is a clone of the exceptional and still enigmatic SN~2008ha \citep{foley2016}. Thus, it enters in the debate concerning the nature and origins of these events: core-collapse or thermonuclear explosions~? \citep{valenti2009,foley2009,moriya2010,foley2016}. Recently, there have been some explorations of explosion scenarios that can produce rapidly-evolving faint transients, even fainter than the faintest SNe~Iax. These include the outcomes of stripped-envelope electron-capture SNe \citep[ECSNe,][]{moriya2016}, as well as a double-degenerate scenario consisting of the merger of a C/O and a O/Ne WD \citep{kashyap2018}. However, to date no spectral modelling for ECSNe \citep{moriya2016} is available and the published synthetic light curves do not cover a wide range in peak luminosity. Moreover, while the light curve properties and ejecta velocities of the faintest SNe~Iax may appear to be somewhat consistent with the stripped-envelope ECSNe model, we stress that the post maximum NIR spectroscopic sequence, obtained by \cite{stritzinger14} for SN 2010ae as well as for the brighter SN 2014ck in \cite{tomasella16} and, earlier, for SN~2005hk in \cite{kromer2013}, indicates the presence of \ion{Co}{ii} and hence provides a stronger link to thermonuclear explosions. Despite the lack of NIR spectra for SN~2019gsc, as well as for SN 2008ha, the striking spectral similarities in the visual-wavelength range with both the faint SN~2010ae, and the brighter SNe 2002cx and 2014ck (when comparing spectra at different epochs, cf. Fig.~\ref{fig:three_panels}), suggest a similar explosion mechanism for the broad class of SNe~Iax.

We stress that the spectroscopic evolution of SN~2019gsc is unusually rapid, as strong forbidden iron and calcium emission lines have clearly emerged in the spectrum taken on +54~d. Equally strong at this epoch are the permitted \ion{Ca}{ii} NIR triplet features. Our late-time spectrum perfectly matches the similar-phase (+44~d) spectrum of SN~2008ha, while SN~2010ae, and other brighter SNe Iax, exhibit calcium lines at a later phase. At any rate, the fainter SNe 2019gsc and 2008ha are spectroscopically similar, if compared at different phases, to the equally faint SN~2010ae and to brighter objects such as SNe 2002cx and 2014ck, plotted in Fig.~\ref{fig:three_panels}. More explicitly, the spectra of SNe 2008ha and 2019gsc $\gtrsim$ 1 month post maximum share the general shape (mostly because of the features content) as other SNe Iax at phases $\gtrsim  5-6$ months post maximum. 

However, as highlighted by \cite{foley2016}, there is also significant diversity in the late-time spectra of SN Iax class, especially in the strength, width and line shifts of forbidden emissions, such as [\ion{Ca}{ii}] $\lambda\lambda$7291, 7324, [\ion{Fe}{ii}] $\lambda$7155, and [\ion{Ni}{ii}] $\lambda$7378. 
\cite{foley2016} proposed a two-component model were the photosphere, P~Cygni, and narrow forbidden lines are produced by the wind $-$ which could be caused by the bound remnant of the progenitor WD and driven by the $^{56}$Ni left in such a remnant $-$, 
while the broad forbidden lines come from the SN ejecta. In this perspective, SN~2019gsc belongs to a low-velocity, narrow-[\ion{Ca}{ii}], [\ion{Ni}{ii}]-poor group (cf. Fig.~\ref{fig:three_panels}), as well as SN~2008ha, but also the bright prototype Iax SN 2002cx \citep[][cf. their Section 3]{foley2016}. The deflagration of a $M_{\textit{Ch}}$ hybrid C/O/Ne WD, leaving behind a bound remnant \citep{kromer2015} might produce faint transients (i.e., having peak absolute $B$ magnitudes in the range $-13.2$ to $-14.4$ mag) with observational characteristics similar to both SNe 2008ha and 2019gsc, though there are shortcomings, for example the spectro-photometric evolution of the model is too fast. A possible alternative is the merger of a O/Ne and C/O WD (double-degenerate channel) that yields a failed detonation and produces a very faint, rapidly fading transient (actually fainter than SNe 2008ha, 2010ae and 2019gsc) with a small $^{56}$Ni and ejecta mass \citep[see][for details]{kashyap2018}. In support of the single-degenerate SN Iax progenitor scenario come the recent works of \cite{vennes2017} and \cite{raddi2019}, with the discovery of a handful of high proper motion, low-mass Galactic WDs traveling at a velocity greater than the Milky Way escape velocity and whose peculiar atmospheres are dominated by IMEs. The authors argue that these inflated and IME contaminated WDs are the partially burnt remnants of SNe~Iax, ejected from a binary SN progenitor with a high kick velocity.
 
 Overall, the balance of evidence seems to suggest that SNe~Iax are associated with the partial disruption of their progenitor stars, even if few works seem to suggest that the brightest members of the class may be linked to the full disruption of a Chandrasekhar mass WD \citep[e.g.,][]{sahu2008,stritzinger15}. 
Anyway, further study of both exceptionally faint and bright SNe~Iax is critical to constrain the progenitors and the explosion mechanisms of this class of transients.

\section*{Acknowledgements}

Special thanks to the GTC Director and Support Staff for executing observations under director discretionary program DDT-GTC2019-129 (P.I. N. Elias-Rosa). 
M.S. is supported by a generous grant (13261) from the VILLUM FONDEN and a project grant from the Independent Research Fund Denmark (IRFD; 8021-00170B). K.M. acknowledges support from H2020 ERC grant no. 758638.
Based on observations made with the Gran Telescopio Canarias (GTC), installed in the Spanish Observatorio del Roque de los Muchachos of the Instituto de Astrofísica de Canarias, in the island of La Palma.
This work is partly based on NUTS2 observations made
with the NOT (program 59-506), operated by the NOT Scientific Association
at the Observatorio del Roque de los Muchachos, La
Palma, Spain, of IAC. ALFOSC is provided by IAA
under a joint agreement with the University of Copenhagen and NOTSA. NUTS2 is funded in part by the Instrument Center for Danish Astrophysics (IDA). 
The classification spectrum of SN 2019gsc was obtained with the Nordic Optical Telescope (program 59-509), operated by the Nordic Optical Telescope Scientific Association at the Observatorio del Roque de los Muchachos, La Palma, Spain, of the Instituto de Astrof\'isica de Canarias, and was kindly made available by S. Srivastav, D. Malesani, G. Leloudas.  
This work made use of ZTF data. We thank ZTF for access to this valuable public data stream and {\it Lasair} (https://lasair.roe.ac.uk/) which provides a broker system for users to access, visualise and extract science data. Finally, we thank the anonymous Reviewer for the careful revision of the manuscript and the detailed and helpful feedback, which lead to a significant improvement of the manuscript. 





\begin{thebibliography}{99}

\bibitem[\protect\citeauthoryear{Aguado et al.}{2019}]{aguado2019} Aguado D.~S., et al., 2019, ApJS, 240, 23
\bibitem[\protect\citeauthoryear{Arnett}{1982}]{arnett1982} Arnett W.~D., 1982, ApJ, 253, 785

\bibitem[\protect\citeauthoryear{Asplund et al.}{2009}]{asplund2009} Asplund M., Grevesse N., Sauval A.~J., Scott P., 2009, ARA\&A, 47, 481

\bibitem[\protect\citeauthoryear{Branch et al.}{2004}]{branch2004} Branch D., Baron E., Thomas R.~C., Kasen D., Li W., Filippenko A.~V., 2004, PASP, 116, 903

\bibitem[\protect\citeauthoryear{Brown et al.}{2009}]{brown2009} Brown P.~J., et al., 2009, AJ, 137, 4517

\bibitem[\protect\citeauthoryear{Cardelli, Clayton \& Mathis}{1989}]{cardelli1989} Cardelli J.~A., Clayton G.~C., Mathis J.~S., 1989, ApJ, 345, 245
\bibitem[\protect\citeauthoryear{Chomiuk et al.}{2016}]{chomiuk2016} Chomiuk L., et al., 2016, ApJ, 821, 119
\bibitem[\protect\citeauthoryear{Chornock, et al.}{2006}]{chornock2006} Chornock R., Filippenko A.~V., Branch D., Foley R.~J., Jha S., Li W., 2006, PASP, 118, 722
\bibitem[\protect\citeauthoryear{Dessart \& Hillier}{2015}]{dessart2015} Dessart L., Hillier D.~J., 2015, MNRAS, 447, 1370

\bibitem[\protect\citeauthoryear{Dopita et al.}{2016}]{dopita2016} Dopita M.~A., Kewley L.~J., Sutherland R.~S., Nicholls D.~C., 2016, Ap\&SS, 361, 61

\bibitem[\protect\citeauthoryear{Filippenko et al.}{2007}]{filippenko2007} Filippenko A.~V., Foley R.~J., Silverman J.~M., Chornock R., Li W., Blondin S., Matheson T., 2007, CBET, 926, 1

\bibitem[\protect\citeauthoryear{Fink et al.}{2014}]{fink2014} Fink M., et al., 2014, MNRAS, 438, 1762

\bibitem[\protect\citeauthoryear{Firth, et al.}{2015}]{firth2015} Firth R.~E., et al., 2015, MNRAS, 446, 3895


\bibitem[\protect\citeauthoryear{Folatelli et al.}{2012}]{folatelli2012} Folatelli G., et al., 2012, ApJ, 745, 74

\bibitem[\protect\citeauthoryear{Foley et al.}{2009}]{foley2009} Foley R.~J., et al., 2009, AJ, 138, 376

\bibitem[\protect\citeauthoryear{Foley et al.}{2013}]{foley2013} Foley R.~J., et al., 2013, ApJ, 767, 57

\bibitem[\protect\citeauthoryear{Foley et al.}{2016}]{foley2016} Foley R.~J., Jha S.~W., Pan Y.-C., Zheng W.~K., Bildsten L., Filippenko A.~V., Kasen D., 2016, MNRAS, 461, 433


\bibitem[\protect\citeauthoryear{Harutyunyan et al.}{2008}]{avik2008} Harutyunyan A.~H., et al., 2008, \aap, 488, 383

\bibitem[\protect\citeauthoryear{Hoeflich, Khokhlov \& Mueller}{1992}]{hoeflich1992} Hoeflich P., Khokhlov A., Mueller E., 1992, A\&A, 259, 549

\bibitem[\protect\citeauthoryear{Huchtmeier et al.}{2008}]{h2008} Huchtmeier W.~K., Petrosian A., Gopal-Krishna, McLean B., Kunth D., 2008, \aap, 492, 367

\bibitem[\protect\citeauthoryear{Jacobson-Gal{\'a}n et al.}{2019}]{jacobson2019} Jacobson-Gal{\'a}n W.~V., et al., 2019, MNRAS, 487, 2538

\bibitem[\protect\citeauthoryear{Jha et al.}{2006}]{jha2006} Jha S., et al., 2006, AJ, 132, 189

\bibitem[\protect\citeauthoryear{Jha}{2017}]{jha2017} Jha S.~W., 2017, hsn..book,  375, hsn..book

\bibitem[\protect\citeauthoryear{Jordan et al.}{2012}]{jordan2012} Jordan G.~C., Perets H.~B., Fisher R.~T., van Rossum D.~R., 2012, ApJL, 761, L23

\bibitem[\protect\citeauthoryear{Kashyap et al.}{2018}]{kashyap2018} Kashyap R., Haque T., Lor{\'e}n-Aguilar P., Garc{\'\i}a-Berro E., Fisher R., 2018, ApJ, 869, 140

\bibitem[\protect\citeauthoryear{Kerzendorf \& Sim}{2014}]{kerze2014} Kerzendorf W.~E., Sim S.~A., 2014, MNRAS, 440, 38

\bibitem[\protect\citeauthoryear{Kromer et al.}{2013}]{kromer2013} Kromer M., et al., 2013, MNRAS, 429, 2287

\bibitem[\protect\citeauthoryear{Kromer et al.}{2015}]{kromer2015} Kromer M., et al., 2015, MNRAS, 450, 3045

\bibitem[\protect\citeauthoryear{Kulkarni}{2018}]{kulkarni2018} Kulkarni S.~R., 2018, ATel, 11266, 1

\bibitem[\protect\citeauthoryear{Leloudas et al.}{2019}]{leloudas2019} Leloudas G., et al., 2019, TNSAN, 25, 1

\bibitem[\protect\citeauthoryear{Li et al.}{2003}]{li2003} Li W., et al., 2003, PASP, 115, 453

\bibitem[\protect\citeauthoryear{Liu et al.}{2010}]{liu2010} Liu W.-M., Chen W.-C., Wang B., Han Z.~W., 2010, A\&A, 523, A3

\bibitem[\protect\citeauthoryear{Lyman et al.}{2018}]{lyman2018} Lyman J.~D., Taddia, F., Stritzinger, M., et al., 2018, MNRAS, 473, 1359

\bibitem[\protect\citeauthoryear{Magee, et al.}{2016}]{magee2016} Magee M.~R., et al., 2016, A\&A, 589, A89

\bibitem[\protect\citeauthoryear{Magee et al.}{2019}]{magee2019} Magee M.~R., Sim S.~A., Kotak R., Maguire K., Boyle A., 2019, A\&A, 622, A102

\bibitem[\protect\citeauthoryear{Marino et al.}{2013}]{marino2013} Marino R.~A., et al., 2013, A\&A, 559, A114

\bibitem[\protect\citeauthoryear{McCully et al.}{2014}]{mccully2014} McCully C., et al., 2014, Natur, 512, 54

\bibitem[\protect\citeauthoryear{Moriya et al.}{2010}]{moriya2010} Moriya T., et al., 2010, ApJ, 719, 1445

\bibitem[\protect\citeauthoryear{Moriya \& Eldridge}{2016}]{moriya2016} Moriya T.~J., Eldridge J.~J., 2016, MNRAS, 461, 2155

\bibitem[\protect\citeauthoryear{Mould et al.}{2000}]{mould2000} Mould J.~R., et al., 2000, ApJ, 529, 786

\bibitem[\protect\citeauthoryear{Nomoto, Kamiya \& Nakasato}{2013}]{nomoto2013} Nomoto K., Kamiya Y., Nakasato N., 2013, IAUS,  253, IAUS..281

\bibitem[\protect\citeauthoryear{Nonaka, et al.}{2012}]{nonaka2012} Nonaka A., Aspden A.~J., Zingale M., Almgren A.~S., Bell J.~B., Woosley S.~E., 2012, ApJ, 745, 73


\bibitem[\protect\citeauthoryear{Parrent et al.}{2011}]{parrent2011} Parrent J.~T., et al., 2011, ApJ, 732, 30


\bibitem[\protect\citeauthoryear{Pettini \& Pagel}{2004}]{pettini2004} Pettini M., Pagel B.~E.~J., 2004, MNRAS, 348, L59

\bibitem[\protect\citeauthoryear{Phillips}{1993}]{phillips1993} Phillips M.~M., 1993, ApJL, 413, L105

\bibitem[\protect\citeauthoryear{Phillips et al.}{2007}]{phillips2007} Phillips M.~M., et al., 2007, PASP, 119, 360
\bibitem[\protect\citeauthoryear{Piro \& Nakar}{2014}]{piro2014} Piro A.~L., Nakar E., 2014, ApJ, 784, 85

\bibitem[\protect\citeauthoryear{Raddi et al.}{2019}]{raddi2019} Raddi R., et al., 2019, MNRAS, 489, 1489

\bibitem[\protect\citeauthoryear{Riess et al.}{1999}]{riess1999} Riess A.~G., et al., 1999, AJ, 118, 2675

\bibitem[\protect\citeauthoryear{Riess et al.}{2016}]{riess2016} Riess A.~G., et al., 2016, ApJ, 826, 56

\bibitem[\protect\citeauthoryear{Sahu et al.}{2008}]{sahu2008} Sahu D.~K., et al., 2008, ApJ, 680, 580

\bibitem[\protect\citeauthoryear{Schlafly \& Finkbeiner}{2011}]{schlafly2011} Schlafly E.~F., Finkbeiner D.~P., 2011, ApJ, 737, 103
\bibitem[\protect\citeauthoryear{Smartt, et al.}{2019}]{smartt2019} Smartt S.~J., et al., 2019, TNSAN, 23, 1
\bibitem[\protect\citeauthoryear{Smith, et al.}{2019}]{smith2019} Smith K.~W., et al., 2019, RNAAS, 3, 26


\bibitem[\protect\citeauthoryear{Srivastav, et al.}{2020}]{srivastav2020} Srivastav S., et al., 2020, ApJL, 892, L24


\bibitem[\protect\citeauthoryear{Stritzinger \& Leibundgut}{2005}]{stritzinger2005} Stritzinger M., Leibundgut B., 2005, A\&A, 431, 423

\bibitem[Stritzinger et al.(2014)]{stritzinger14} Stritzinger M.~D., et al., 2014, \aap, 561, A146

\bibitem[Stritzinger et al.(2015)]{stritzinger15} Stritzinger M.~D., et al., 2015, \aap, 573, A2
\bibitem[\protect\citeauthoryear{Taubenberger}{2017}]{taub2017} Taubenberger S., 2017, hsn..book,  317, hsn..book

\bibitem[\protect\citeauthoryear{Taubenberger, et al.}{2019}]{taub2019} Taubenberger S., et al., 2019, MNRAS, 488, 5473



\bibitem[Tomasella et al.(2016)]{tomasella16}
Tomasella L., et al., 2016, \mnras, 459, 1018



\bibitem[\protect\citeauthoryear{Tonry, et al.}{2018}]{tonry2018} Tonry J.~L., et al., 2018, PASP, 130, 064505


\bibitem[\protect\citeauthoryear{Valenti et al.}{2009}]{valenti2009} Valenti S., et al., 2009, Natur, 459, 674

\bibitem[\protect\citeauthoryear{Vennes et al.}{2017}]{vennes2017} Vennes S., Nemeth P., Kawka A., Thorstensen J.~R., Khalack V., Ferrario L., Alper E.~H., 2017, Sci, 357, 680




\end{thebibliography}








\bsp	
\label{lastpage}
\end{document}